\documentclass{aa}  
\usepackage{cleveref}

\usepackage{graphicx}
\usepackage{txfonts}
\usepackage{multirow}
\begin{document}

   \title{Discovery of a giant and luminous Ly$\alpha$+C{\sc iv}+He{\sc ii} nebula at $z = 3.326$ with extreme emission line ratios}

\titlerunning{Discovery of a giant and luminous Ly$\alpha$+C{\sc iv}+He{\sc ii} nebula at $z = 3.326$}

   \author{R. Marques-Chaves\inst{1,2,3}\fnmsep\thanks{email: rmarques@cab.inta-csic.es}
    \and I. P\'{e}rez-Fournon\inst{1,3}
    \and M. Villar-Mart\'in\inst{2}
    \and R. Gavazzi\inst{4}
    \and D. Riechers\inst{5}
    \and D. Rigopoulou\inst{6}
    \and J. Wardlow\inst{7}    
    \and A. Cabrera-Lavers\inst{1,8}
    \and D. L. Clements\inst{9}
    \and L. Colina\inst{2}
    \and A. Cooray\inst{10}
    \and D. Farrah\inst{11, 12}
    \and R. J. Ivison\inst{13, 14}
    \and C. Jim\'{e}nez-\'{A}ngel\inst{1,3}
    \and P. Mart\'\i nez-Navajas\inst{1,3}
    \and H. Nayyeri\inst{10}
    \and S. Oliver\inst{15}
    \and A. Omont\inst{4}
    \and D. Scott\inst{16}
    \and Y. Shu\inst{17}
          }

   \institute{
    $^{1}$
    Instituto de Astrof\'\i sica de Canarias, C/V\'\i a L\'actea, s/n, E-38205 San Crist\'obal de La Laguna, Tenerife, Spain\\
    $^{2}$
    Centro de Astrobiolog\'ia (CSIC-INTA), Carretera de Ajalvir, 28850 Torrej\'on de Ardoz, Madrid, Spain\\
    $^{3}$
    Universidad de La Laguna, Dpto. Astrof\'\i sica, E-38206 San Crist\'obal de La Laguna, Tenerife, Spain\\
    $^{4}$
    Institut d'Astrophysique de Paris, UMR7095 CNRS \& Sorbonne Universit\'e (UPMC), F-75014 Paris, France\\
    $^{5}$
    Astronomy Department, Cornell University, Ithaca, NY 14853, USA\\
    $^{6}$
    Astrophysics, Department of Physics, University of Oxford, Keble Road, Oxford, OX1 3RH, UK\\
    $^{7}$
    Department of Physics, Lancaster University, Lancaster, LA1 4YB, UK\\
    $^{8}$
    GRANTECAN, Cuesta de San Jos\'{e} s/n, E-38712, Bre\~{n}a Baja, La Palma, Spain\\
    $^{9}$
    Astrophysics Group, Imperial College London, Blackett Laboratory, Prince Consort Road, London SW7 2AZ, UK\\
    $^{10}$
    Department of Physics and Astronomy, University of California, Irvine, CA 92697, USA\\
    $^{11}$
    Department of Physics and Astronomy, University of Hawaii, 2505 Correa Road, Honolulu, HI 96822, USA\\
    $^{12}$
    Institute for Astronomy, 2680 Woodlawn Drive, University of Hawaii, Honolulu, HI 96822, USA\\
    $^{13}$
    European Southern Observatory, Karl-Schwarzschild-Str. 2, D-85748 Garching, Germany\\
    $^{14}$    
    Institute for Astronomy, University of Edinburgh, Royal Observatory, Blackford Hill, Edinburgh EH9 3HJ, UK\\
    $^{15}$
    Astronomy Centre, Department of Physics and Astronomy, University of Sussex, Brighton BN1 9QH, UK\\
    $^{16}$
    Department of Physics and Astronomy, University of British Columbia, 6224 Agricultural Road, Vancouver, BC V6T 1Z1, Canada\\ 
    $^{17}$
    Institute of Astronomy, University of Cambridge, Madingley Road, Cambridge CB3 0HA, UK
    }

   \date{Received:; accepted:}

% \abstract{}{}{}{}{} 
% 5 {} token are mandatory
 
  \abstract
{We present the discovery of HLock01-LAB, a luminous and large Ly$\alpha$ nebula at $z=3.326$. Medium-band imaging and long-slit spectroscopic observations with the Gran Telescopio Canarias reveal extended emission in the Ly$\alpha$~1215\AA, C~{\sc iv}~1550\AA, and He~{\sc ii}~1640\AA{ }lines over $\sim 100$~kpc, and a total luminosity $L_{\rm Ly\alpha} = (6.4 \pm 0.1) \times 10^{44}$ erg s$^{-1}$. HLock01-LAB presents an elongated morphology aligned with two faint radio sources contained within the central $\sim 8$~kpc of the nebula. The radio structures are consistent to be faint radio jets or lobes of a central galaxy, whose spectrum shows nebular emission characteristic of a type-II active galactic nucleus (AGN). The continuum emission of the AGN at short wavelengths is, however, likely dominated by stellar emission of the host galaxy, for which we derive a stellar mass $M_{*} \simeq 2.3 \times 10^{11}$~M$_{\odot}$. Our kinematic analysis shows that the ionized gas is perturbed almost exclusively in the inner region between the radio structures, probably as a consequence of jet-gas interactions, whereas in the outer regions the ionized gas appears more quiescent. The detection of extended emission in C~{\sc iv} and C~{\sc iii]} indicates that the gas within the nebula is not primordial. Feedback may have enriched the halo at at least 50 kpc from the nuclear region. Using rest-frame UV emission-line diagnostics, we find that the gas in the nebula is likely heated by the AGN. Nevertheless, at the center of the nebula we find extreme emission line ratios of Ly$\alpha$/C{\sc iv} $\sim 60$ and Ly$\alpha$/He{\sc ii} $\sim 80$, one of the highest values measured to date, and well above the standard values of photoionization models (Ly$\alpha$/He{\sc ii} $\sim 30$ for case B photoionization). Our data suggest that jet-induced shocks are likely responsible for the increase of the electron temperature and, thus, the observed Ly$\alpha$ enhancement in the center of the nebula. This scenario is further supported by the presence of radio structures and perturbed kinematics in this region. 
The large Ly$\alpha$ luminosity in HLock01-LAB is likely due to a combination of AGN photoionization and jet-induced shocks, highlighting the diversity of sources of energy powering Ly$\alpha$ nebulae.
Future follow-up observations of HLock01-LAB will help in revealing in more detail the excitation conditions of the gas induced by jets and investigate the underlying cooling and feedback processes in this unique object.

  }


   \maketitle
%
%________________________________________________________________

\section{Introduction}

Extended regions of Ly$\alpha$ emission were initially discovered around high redshift powerful radio sources \citep{chambers1990, heckman1991b, heckman1991a}. Later on, dedicated narrow-band imaging surveys have discovered similar Ly$\alpha$ nebulae in overdense regions with no clear association to radio galaxies, also called as Ly$\alpha$ blobs \citep[LABs, e.g.,][]{francis1996, fynbo1999, keel1999, steidel2000}. 

These spectacular objects are characterized by large Ly$\alpha$ luminosities ($\sim 10^{43-44}$ erg s$^{-1}$) with sizes of up to hundreds kpc \citep[e.g.,][]{matsuda2004} or more \citep{cantalupo2014, arrigoni2018, cai2018, arrigoni2019}. 
Ly$\alpha$ nebulae have been found associated with a diverse population of galaxies, from powerful high-$z$ radio galaxies \citep[HzRGs; e.g.,][]{chambers1990, kurk2002, reuland2003, villar2003, venemans2007, villar2007}, quasi stellar objects \citep[QSOs; e.g.,][]{heckman1991b, heckman1991a, bunker2003, weidinger2004, christensen2006, cantalupo2014, borisova2016, arrigoni2018}, Lyman-break galaxies \citep[LBGs; e.g.,][]{matsuda2004}, and sub-millimetre galaxies \citep[SMGs;  e.g.,][]{ivison1998, chapman2001, geach2005, matsuda2007, geach2014, oteo2017a, li2019}. 
Many others have been found without any clear galactic counterpart \citep[e.g.,][]{nilsson2006}, although deep data have revealed that the majority of them are associated with highly obscured active galactic nuclei \citep[type-II AGNs; e.g.,][]{dey2005, geach2009, bridge2013, overzier2013, hennawi2015, ao2017}. 
These extended regions of Ly$\alpha$ emission are expected to occupy the densest dark matter regions of the Universe, tracing large-scale mass overdensities \citep[e.g.,][]{steidel2000, matsuda2004, prescott2008, saito2015, cai2017}. 

There are several possible explanations for the origin of circumgalactic Ly$\alpha$ emission. These include the photoinization radiation from strong ultra-violet (UV) ionizing sources \citep[e.g.,][]{cantalupo2005, geach2009, kollmeier2010}, radiation from shock-heated gas powered by relativistic winds or jets 
\citep[e.g.][]{taniguchi2000, allen2008}, resonant scattering of Ly$\alpha$ \citep[e.g.,][]{villar1996, hayes2011, cantalupo2014}, or cooling radiation when the gas falls towards galaxies \citep[e.g.,][]{fardal2001, dijkstra2009}. 
However, investigating the physical process powering the emission in these Ly$\alpha$ nebulae is challenging, in particular if only the Ly$\alpha$ line is available.

Although typically less luminous than Ly$\alpha$, the detection of other UV emission lines, such as N~{\sc v}~1238,1240\AA, C~{\sc iv} 1548,1550\AA, He~{\sc ii} 1640\AA, or C~{\sc iii]}~1906,1908\AA{ }(hereafter N~{\sc v}, C~{\sc iv}, He~{\sc ii}, and C~{\sc iii]}, respectively) among others, may provide us key information on the properties of the gas and can help in disentangle the main physical process powering these nebulae \citep[e.g.,][]{villar1997, villar2007, arrigoni2015a, arrigoni2015b, feltre2016, nakajima2017, humphrey2019}. 
For example, He~{\sc ii} is a non-resonant (and recombination) line and it is possible to test whether or not Ly$\alpha$ photons are being resonantly scattered, by comparing the morphology and kinematics of both He~{\sc ii} and Ly$\alpha$ extended emission. In addition, the C~{\sc iii]} and C~{\sc iv} metallic lines are useful to constrain the size at which the halo is metal-enriched, and to investigate the intensity and hardness of the ionizing sources. 
Extended emission in these lines have been detected on scales up to $100$~kpc in some HzRGs, showing relatively high surface brightness with perturbed kinematics (full width at half maximum $\rm FWHM \gtrsim 1000$~km~s$^{-1}$) confined by (and aligned with) the radio structures, as further evidence for jet-gas interactions \citep[e.g.,][]{villar2003, humphrey2006, morais2017}.
However, these lines appear to be very faint and difficult to observe in Ly$\alpha$ nebulae associated with  sources other than HzRGs \citep[e.g.,][]{arrigoni2015b, borisova2016}, with only a few detections reported so far \citep[e.g.,][]{dey2005, prescott2009, prescott2013, caminha2016, cai2017, cantalupo2019, marino2019}.

In this paper, we present the discovery and first characterization of the observed emission line properties and the interpretation using photoionization and shock models of a new Ly$\alpha$ nebula at $z = 3.3$, nicknamed ``HLock01-LAB'' hereafter. 
The nebula is powered by a central AGN with two faint and compact radio structures, yet it is one of the most luminous Ly$\alpha$ nebulae known, in contrast with the general idea that Ly$\alpha$ halos around powerful HzRGs show statistically larger Ly$\alpha$ luminosities with respect to fainter HzRGs \citep[e.g.,][]{heckman1991b, miley2006, saxena2019}.
HLock01-LAB shows extended emission over $\sim 100$~kpc in the Ly$\alpha$, C~{\sc iv}, and He~{\sc ii} lines, and presents one of the highest values of Ly$\alpha$/C~{\sc iv} and Ly$\alpha$/He~{\sc ii} measured to date.
It is located close in projection ($\sim 15^{\prime \prime}$), but physically unrelated to the strong gravitationally lensed {\it Herschel} galaxy HLock01 at $z\simeq 2.96$ already discussed in several works \citep{conley2011, riechers2011, gavazzi2011, scott2011, bussmann2013, calanog2014, marques2018, rigopoulou2018}. 

The paper is structured as follows. The discovery and follow-up observations are presented in Section \ref{sec:2}. The analysis and discussion of imaging and spectroscopic data are presented in Section \ref{sec:3}. In Section \ref{sec:4} we compare the properties of HLock01-LAB with those from other Ly$\alpha$ nebulae, and, finally, in Section \ref{sec:5} we present the summary of our main findings.
Throughout this work, a cosmology with $\Omega_{\rm m} = 0.274$, $\Omega_{\Lambda} = 0.726$, and $H_{0} = 70$ km s$^{-1}$ Mpc$^{-1}$ is adopted. At $z\sim 3.3$, $1^{\prime \prime}$ corresponds to $7.66$~kpc. All quoted magnitudes are in the AB system.

\section{Discovery and Follow-up Observations}\label{sec:2}

\subsection{Serendipitous discovery}

\begin{figure}[h!]
\centering
$\begin{array}{rl}
    \includegraphics[width=0.47\textwidth]{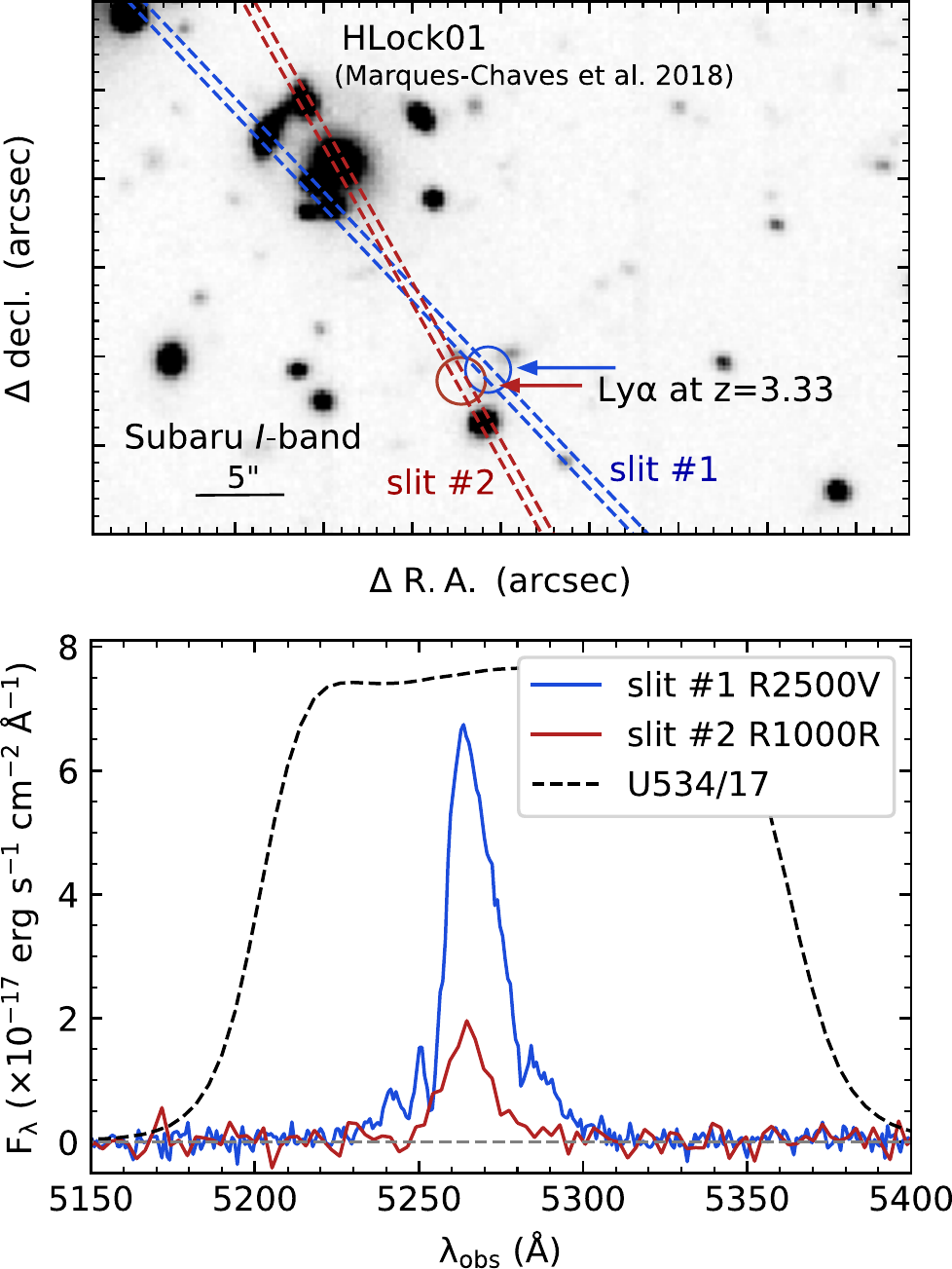}
\end{array}$
\caption{Serendipitous discovery of a bright Ly$\alpha$ emitting region, HLock01-LAB. Top: Subaru $I$-band imaging showing the locations of the OSIRIS long slits on the sky (dashed lines) used in the analysis of the bright gravitational lens HLock01 in \cite{marques2018} (top left corner). The approximate positions where the bright Ly$\alpha$ emission was serendipitously detected are marked with circles.
Bottom: spectra encompassing the region of the Ly$\alpha$ emission detected in both slits. The location of the two slits is shown in the top panel. The dashed line shows the transmission of the SHARDS U534/17 medium-band filter. \label{fig:4_1}}
\end{figure}

In \cite{marques2018} we reported the serendipitous detection of a bright asymmetric line at $5264$~\AA{ }close to the strong gravitational lensed system HLock01 ($z \simeq 2.96$), consistent with Ly$\alpha$ emission at $z \sim 3.3$. The Ly$\alpha$ line was detected in different spatial positions $\simeq 15^{\prime \prime}$ SW from HLock01 (see Figure \ref{fig:4_1}) in two long-slit spectroscopic observations with the Gran Telescopio Canarias (GTC).
The approximate locations of the peaks of the Ly$\alpha$ emission found in both long-slit spectra (slit \#1 and \#2 in Figure \ref{fig:4_1}) are spatially separated by $\sim 2^{\prime \prime}$.  
The total Ly$\alpha$ flux within the two slits is $\simeq 1.8 \times 10^{-15}$~erg~s$^{-1}$~cm$^{-2}$, which corresponds to a Ly$\alpha$ luminosity of $\simeq 1.9 \times 10^{44}$ erg s$^{-1}$ at $z = 3.33$ (no extinction correction has been applied). 
The observed Ly$\alpha$ luminosity is much higher ($40 \times L_{\rm Ly \alpha}^{*}$) than those found in typical Ly$\alpha$ emitting galaxies (LAEs) at similar redshifts \citep[$L_{\rm Ly \alpha}^{*} \sim 5 \times 10^{42}$ erg s$^{-1}$, e.g.,][]{ouchi2008, sobral2017} or in other exceptionally luminous LAEs \citep[e.g.,][]{ouchi2009, sobral2015, sobral2018b, marques2017}. All this together suggest that the Ly$\alpha$ emission comes from a more extended region, similar to what is found in high-$z$ Ly$\alpha$ nebulae around AGNs. 

\begin{figure*}[h!]
\centering
$\begin{array}{rl}
    \includegraphics[width=0.65\textwidth]{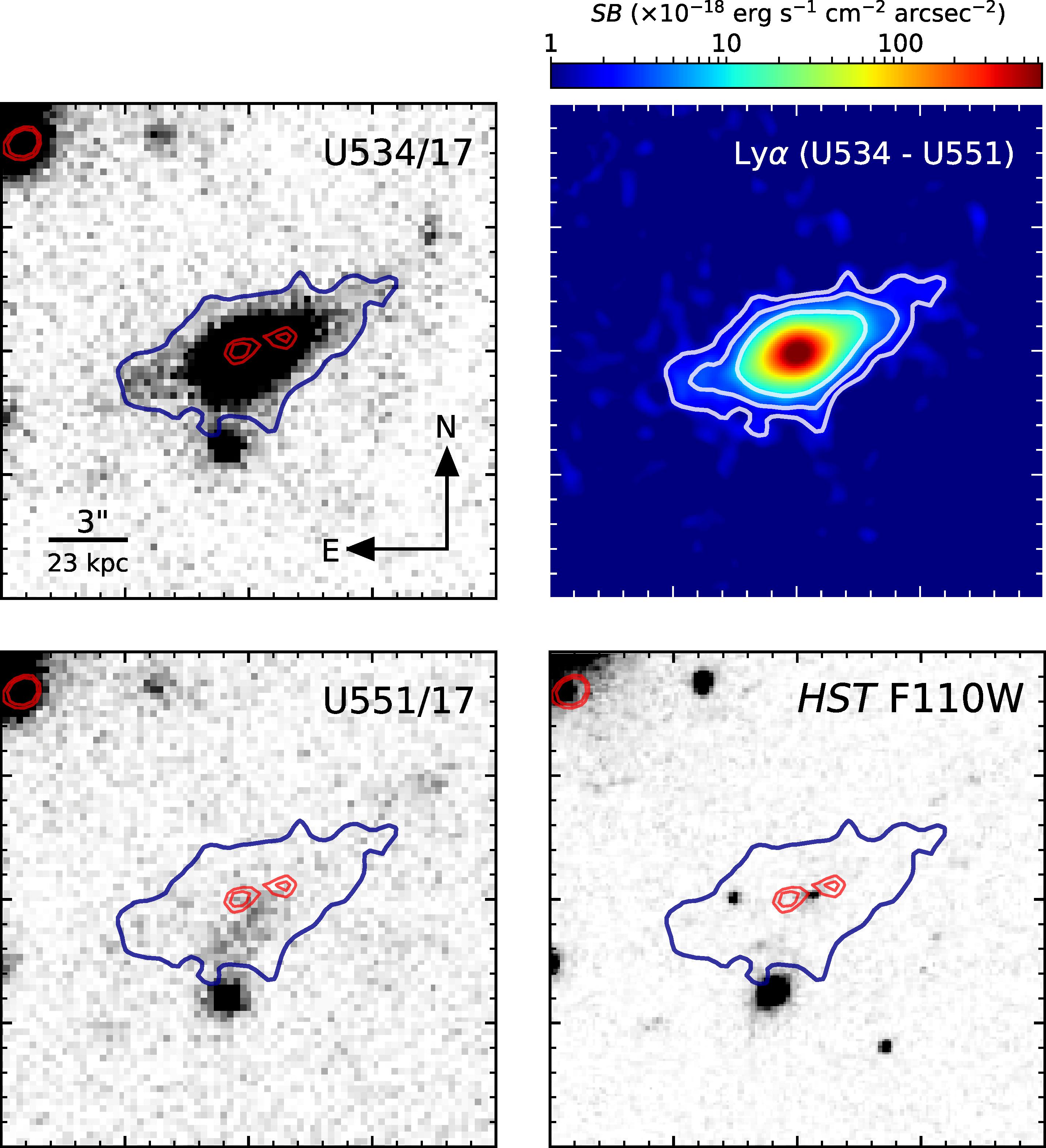}
\end{array}$
\caption{Cutouts images of HLock01-LAB: SHARDS medium-band U534/17 ($\lambda_{\rm cent} \simeq 5300$~\AA, $\rm FWHM \simeq 177$ \AA) and U551/17 ($\lambda_{\rm cent} \simeq 5500$~\AA, $\rm FWHM \simeq 138$ \AA) (upper and bottom left, respectively), continuum-subtracted Ly$\alpha$ emission smoothed with a Gaussian kernel (U534/17-U551/17, upper right), and $HST$ WFC3 F110W (bottom right). The size of the images is $20^{\prime \prime} \times 20^{\prime \prime}$ centered on the Ly$\alpha$ nebula. 
Red and blue contours indicate respectively the radio VLA 1.4 GHz (3 and 4$\sigma$ levels) and the the 3$\sigma$ Ly$\alpha$ emission (5 and 10$\sigma$ levels are also shown in white in the smoothed Ly$\alpha$ image). 
The Ly$\alpha$ emission shows an elongated morphology aligned with the radio sources contained within the central $\simeq 8$~kpc of the nebula. In all panels North is up and East is to the left.
\label{fig:4_2}}
\end{figure*}

\subsection{Ly$\alpha$ imaging with medium-band SHARDS filters}

To understand the origin of the Ly$\alpha$ emission, we use the GTC Optical System for Imaging and low-Intermediate Resolution Integrated Spectroscopy instrument (OSIRIS)\footnote{\url{http://www.gtc.iac.es/instruments/osiris/}} to obtain deep imaging of the Ly$\alpha$ emission. OSIRIS has a field of view of $7.8^{\prime} \times 8.5^{\prime}$ with a plate scale of $0.254^{\prime \prime}$ pixel$^{-1}$. We use the SHARDS  \citep{perez2013}\footnote{\url{https://guaix.fis.ucm.es/~pgperez/SHARDS/}} medium-band filter U534/17, centered at $\lambda_{\rm cent} \simeq 5300$ \AA, with $\rm FWHM \simeq 177$ \AA{ }(see the transmission curve in the lower panel of Figure \ref{fig:4_1}). Additional observations with the consecutive medium-band filter,  U551/17 ($\lambda_{\rm cent} \simeq 5500$ \AA; $\rm FWHM \simeq 138$ \AA), were obtained to perform the continuum subtraction. 
These observations were obtained in service mode in 2017 April 24 in dark conditions as part of the GTC program GTC61-17A (PI: Marques-Chaves). The total exposure time were 3000 and 3750 s for U534/17 and U551/17, respectively, split into 10 individual exposures of 300 and 375 s, respectively, adopting a $5^{\prime \prime}$ dither pattern. 
Individual frames were reduced following standard reduction procedures using {\sc Iraf}.\footnote{\url{http://iraf.noao.edu/}} 
These include subtraction of the bias and further correction of the flat-field using skyflats.
The registration and combination of the individual images were done using {\sc Scamp} \citep{bertin2006} and {\sc Swarp} \citep{bertin2010}. 
The image astrometry is determined using GAIA DR2 \citep{gaia2018}, 
yielding a rms $\simeq 0.2^{\prime \prime}$.
The seeing of both final images was $0.9^{\prime \prime}$ FWHM. The depths of the U534/17 and U551/17 images are 25.6 and 25.4 AB (5$\sigma$) for point sources, respectively.
As shown later in Section~\ref{size}, the SHARDS medium-band U534/17 image probes only the highest surface brightness regions of Ly$\alpha$. Therefore, we will use instead the $5^{\prime \prime}$-wide long-slit spectroscopic observations to measure the total Ly$\alpha$ flux of the nebula. Figure \ref{fig:4_2} shows SHARDS medium-band images in the region of HLock01-LAB.

\subsection{Long-slit spectroscopic observations}\label{sec:2.2}

Additional GTC/OSIRIS spectroscopic observations of HLock01-LAB were also obtained. These were carried out in service mode in 2017 May 19 under sub-arsecond seeing conditions ($\simeq 0.7^{\prime \prime} - 0.9^{\prime \prime}$ FWHM).  
We used a $1.5^{\prime \prime}$-wide long slit\footnote{The seeing (FWHM 0.7$^{\prime \prime}$-0.9$^{\prime \prime}$) was narrower than the slit width (1.5$^{\prime \prime}$). The potential impact on the kinematic measurements will be mentioned in Section \ref{kin}.} centered on a bright reference star $\simeq 40^{\prime \prime}$ SE and oriented so as to encompass the brightest region of the Ly$\alpha$ emission. The long slit was aligned along the major axis of the nebula as measured from the Ly$\alpha$ image at a sky position angle 
$\rm PA = 110^{\circ}$ (measured North to East, see Figure \ref{fig:4_3}). The GTC grism R1000R was used, providing a spectral resolution of $\simeq 650 - 500$ km s$^{-1}$ within the wavelength range of $5100 - 10000$~\AA, respectively.
The total exposure time was 4650~s, split into 6 individual exposures of 760~s each.
In addition, we obtained another spectrum with a wider long-slit ($5.0^{\prime \prime}$-wide), in an attempt to measure and calibrate the total flux of the Ly$\alpha$ emission. As shown in the left panel of Figure \ref{fig:4_3}, we do not expect considerable slit losses using the wide long-slit. 
The data were processed with standard {\sc Iraf} tasks. Both 1D spectra were extracted and corrected for the instrumental response using observations of the standard stars G191-B2B and GD 153. Atmospheric extinction and air mass have been taken into account in this correction.

\subsection{Ancillary data}

Since this object lies very close to the HLock01 system, we use the ancillary data available in this field that were already discussed in other works \citep{conley2011, riechers2011, bussmann2013, wardlow2013, marques2018}.
These consist of optical imaging from OSIRIS/GTC ($g$-band), MEGACAM ($r$-band) on the Canada-France-Hawaii Telescope (CFHT), Suprime-Cam ($I$-band) on the Subaru Telescope, and near-infrared (IR) imaging from {\it Hubble Space Telescope} ({\it HST}) Wide Field Camera 3 F110W ($1.1$ $\mu$m), and LIRIS $K_{\rm s}$ filter at $2.2$ $\mu$m on the William Herschel Telescope (WHT). 
We also use mid-IR {\it Spitzer} IRAC and MIPS images and catalogs from the Spitzer Extragalactic Representative Volume Survey \citep[SERVS;][]{mauduit2012} and the \textit{Spitzer} Wide-Area InfraRed Extragalactic survey \citep[SWIRE;][]{lonsdale2003}. Data from the Submillimeter Array (SMA) at $880$~$\mu$m and the Combined Array for Research in Millimeter-wave Astronomy (CARMA) at 3300~$\mu$m are also available, but no positive flux is detected at the position of HLock01-LAB at $5\sigma$ confidence levels of 4.1 and 0.8~mJy, respectively.
In addition, radio data from the Karl G. Jansky Very Large Array (VLA) at 1.4 GHz (with beamsize of $\simeq 1.1^{\prime \prime}$ and $\rm rms = 25\mu Jy$ from the program ID 11A-182; PI: Ivison) are available, and two unresolved radio structures are detected within the nebula with flux densities $0.151 \pm 0.03$~mJy and $0.116 \pm 0.03$~ mJy. Finally, this region has been imaged in 
the X-ray by \textit{Chandra} with a total integration time of 4.7~ks. However, HLock01-LAB is located near the edge of the field-of-view, and it is not detected with an X-ray flux limit of $8.6 \times 10^{-14}$~erg~s$^{-1}$~cm$^{-2}$ (0.5-7.0~keV), corresponding to a luminosity limit of $4.6 \times 10^{42}$~erg~s$^{-1}$ at $z=3.3$ (considering a photon index $\Gamma = 1.7$).

\section{Results and Discussion}\label{sec:3}

\begin{figure*}[h!]
\centering
$\begin{array}{rl}
    \includegraphics[width=0.79\textwidth]{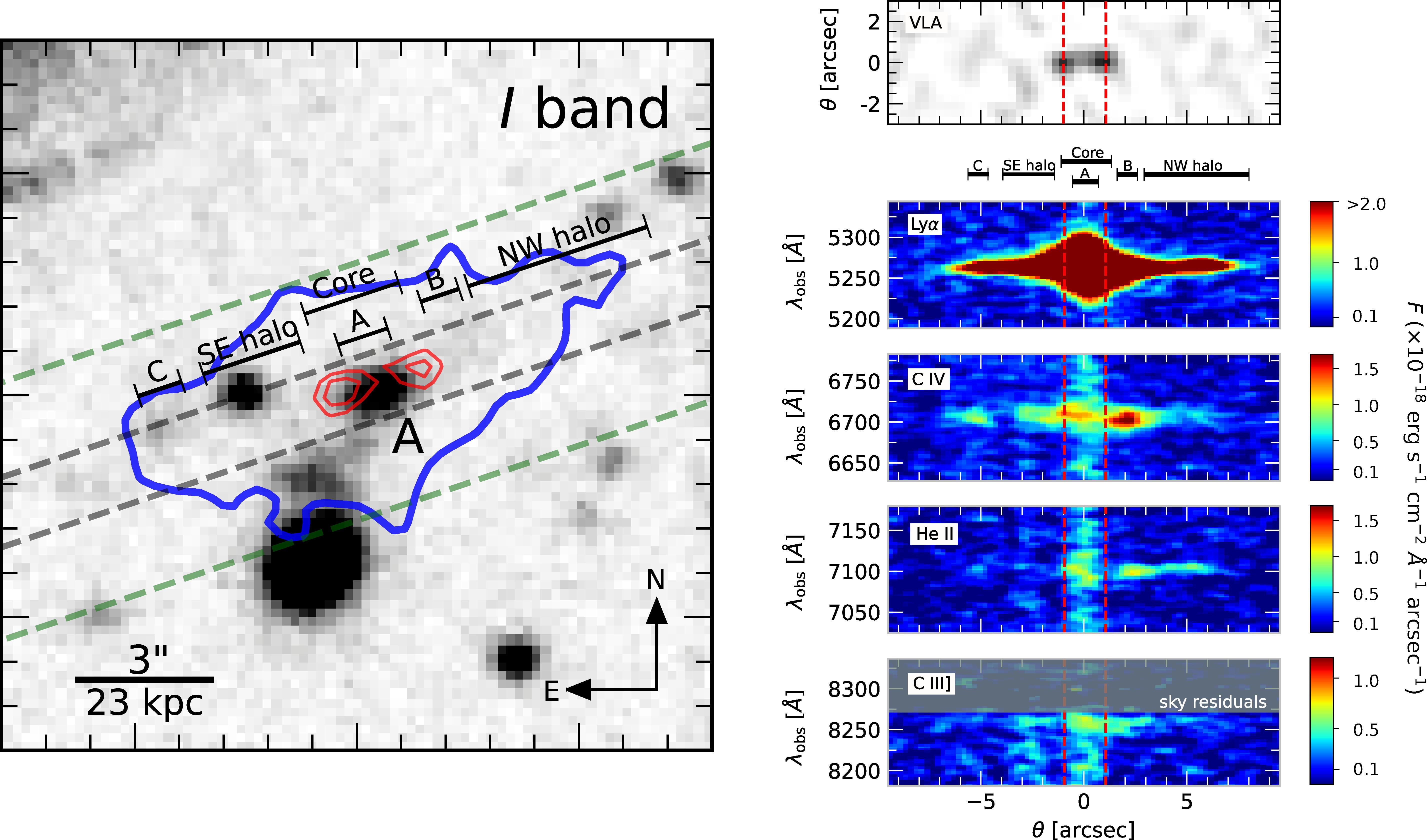}
\end{array}$
\caption{Left: Subaru $I$ band image ($16^{\prime \prime} \times 16^{\prime \prime}$). Blue and red contours represent respectively the $3\sigma$ detection of the Ly$\alpha$ emission and the VLA 1.4 GHz radio emission ($3$ and $4 \sigma$ levels). The central galaxy likely responsible for the Ly$\alpha$ and radio emission is labeled as source ``A''. 
The orientations of the GTC $1.5^{\prime \prime}$- and $5.0^{\prime \prime}$-wide long slits are also plotted with grey and green dashed lines, respectively. 
The approximate location and size of several apertures used to study the emission line spectra of the central source ``A'', the region encompassing the radio emission (core), knots B and C, and the NW and SE extended regions of the nebula are also shown (see text).
Right: VLA 1.4~GHz image (up) and smoothed 2D spectra (down) of Ly$\alpha$, C~{\sc iv}, He~{\sc ii}, and C~{\sc iii]} as indicated (using the 1.5$^{\prime \prime}$-wide long-slit spectra; flux in units of $10^{-18}$~erg~s$^{-1}$~cm$^{-2}$~\AA$^{-1}$~arcsec$^{-1}$, considering a pixel size of 2.55~\AA{ } and $0.25^{\prime \prime}$ in the spectral and spatial dimensions, respectively). Vertical dashed red lines mark the approximate position along the slit of the radio emission. Ly$\alpha$ is detected over $\simeq 15^{\prime \prime}$, which at $z=3.3$ it corresponds to $\simeq 110$ kpc. 
\label{fig:4_3}}
\end{figure*}

\subsection{Projected size and luminosity of the nebula}\label{size}

Figure \ref{fig:4_2} shows the images of HLock01-LAB in the SHARDS medium-bands filters, U534/17 and U551/17 with $\lambda_{\rm cent} \simeq 5300$ and 5500~\AA, respectively. These images probe the Ly$\alpha$ $+$ continuum emission and only continuum emission redward of Ly$\alpha$, respectively. 
The only source in the observed field with a significant excess in the U534/17 image is HLock01-LAB due to the strong Ly$\alpha$ emission (see Figure \ref{fig:4_2}).

The continuum subtracted Ly$\alpha$ image is obtained by estimating and subtracting the continuum emission underlying the U534/17 filter. To do so, we use the continuum emission of HLock01-LAB in the U551/17 filter and assume conservatively a flat UV continuum slope of $\beta = -2$. Even assuming a redder UV slope (e.g., $\beta = -1$), the differences in the continuum emission in U534/17 and U551/17 filters would be negligible ($\Delta m \simeq 0.03$ mag), given the small spectral separation of both medium-band filters, $\simeq 170${ }\AA. We subtracted the flux-matched images after projecting both onto a common world coordinate system. We did not match the point-spread functions (PSF) given that all data were obtained consecutively with similar seeing conditions ($\simeq 0.9^{\prime \prime}$ FWHM). 
In order to accentuate the faintest levels of the extended Ly$\alpha$ emission, we smoothed the resulting continuum-subtracted image using a Gaussian kernel with $\sigma = 1^{\prime \prime}$.
The Ly$\alpha$ nebula (upper right panel in Figure \ref{fig:4_2}) shows an elongated morphology with an orientation of $\sim 110^{\circ}$ (measured North to East) and extends over $\simeq 11^{\prime \prime}$ (or $\simeq 85$~kpc at $z=3.33$) within the $3 \sigma$ detection limit.

The OSIRIS spectrum shows Ly$\alpha$ emission detected over a significantly larger region, about $ 15^{\prime \prime}$, which at $z=3.3$  corresponds to $\simeq 110$ kpc. 
The observed Ly$\alpha$ extension should, however, be regarded as a lower limit, since the total throughput of OSIRIS and the R1000R grism is $\approx 7$\% at $\approx 5250$~\AA. The sensitivity level of the GTC spectrum in the studied spectral range is $\simeq 6 \times 10^{-18}$ erg s$^{-1}$ cm$^{-2}$ ($1 \sigma$) over a $ 1.5^{\prime \prime} \times 1.5^{\prime \prime}$ aperture, corresponding roughly to a surface brightness sensitivity flux of $\rm SB \simeq 3 \times 10^{-18}$ erg s${-1}$ cm$^{-2}$ arcsec$^{-2}$. 
This detection limit is insufficient to detect, if present, fainter levels of the Ly$\alpha$ surface brightness at larger scales, similar to those found over $> 200$ kpc around high-$z$ QSOs and type-II AGNs using very deep integral field unit spectroscopic observations \citep[reaching much deeper flux limits, $\rm SB \sim (0.2 - 1.0) \times 10^{-18}$ erg s$^{-1}$ cm$^{-2}$ arcsec$^{-2}$; e.g.,][]{arrigoni2018, cai2018, arrigoni2019}. 

In Figure \ref{fig:4_3} we show 2D GTC spectra encompassing the region of the Ly$\alpha$ emission, as well as C~{\sc iv}, and He~{\sc ii}. 
C~{\sc iv} appears to have the same extension as Ly$\alpha$, although much fainter, whereas He~{\sc ii} emission is highly asymmetric with emission being detected preferentially on the NW side up to a similar extension as C~{\sc iv} and Ly$\alpha$ (although He~{\sc ii} is also detected at $\simeq 5^{\prime \prime}$ SW of source ``A'', in the spatial region labeled as ``C'' in Figure \ref{fig:4_3}). Emission in C~{\sc iii}] is also detected over the central $\sim 40$~kpc of the nebula, but with very low significance ($< 3\sigma$).
The detection of Ly$\alpha$ and C~{\sc iv} with similar extensions as He~{\sc ii}, at least on the NW side of the nebula, suggests that both are emitted by ionized gas and that resonant scattering plays no significant role on the observed sizes. It is also probable that extended He~{\sc ii} emission is also present on the SE side as well (note that He~{\sc ii} is barely detected in the region marked as “C” on the SE side of the nebula), but with surface brightness levels bellow our detection limits. Notice that Ly$\alpha$ scattering is produced by neutral hydrogen, while highly ionized gas is needed to scatter C~{\sc iv}. 

Turning to the total Ly$\alpha$ luminosity, since the SHARDS medium-band filter only probes the highest surface brightness regions, we use the wide ($5^{\prime \prime}$, displayed with green dashed lines in Figure \ref{fig:4_3}) GTC long-slit spectrum to measure the total Ly$\alpha$ flux. 
Using a large aperture of $15^{\prime \prime}$ along the spatial direction we measure a total Ly$\alpha$ flux $F_{\rm Ly \alpha} = (6.04 \pm 0.08) \times 10^{-15}$ erg s$^{-1}$ cm$^{-2}$, which at $z=3.3$ corresponds to a luminosity $L_{\rm Ly \alpha} = (6.41 \pm 0.08) \times 10^{44}$ erg s$^{-1}$ (without any dust correction).

It is worth noting that HLock01-LAB is located close in projection ($\sim 15^{\prime \prime}$ SW) to the gravitationally lensed system HLock01 at $z\simeq 2.96$ \citep{conley2011, gavazzi2011, riechers2011}. HLock01 is magnified by a group of galaxies with spectroscopic redshifts of $\simeq 0.64$ \citep{marques2018}. However, at $\sim 15^{\prime \prime}$ of the main deflecting galaxy (labeled as ``G1'' in \citealt{gavazzi2011} and \citealt{marques2018}) we do not expect a large magnification on the observable fluxes in the region of the nebula. Using the lens model presented in \cite{marques2018} and taking in consideration its degeneracy due to the large distance to the main deflector, we estimate an upper limit on the magnification of $\mu \lesssim 1.5$ in the region of HLock01-LAB. Even assuming $\mu = 1.5$, the intrinsic (de-magnified) properties
of HLock01-LAB (the corrected Ly$\alpha$ luminosity and size would be respectively $4.3 \times 10^{44}$ erg s$^{-1}$ and $\simeq 90$~kpc) do not change the main results of this work. Furthermore, we do not expect any differential magnification that would change the emission line ratios or equivalent widths of the lines.

\subsection{The central galaxy ``A''}

The deep Subaru $I$-band image ($\sim$1800~\AA{ }rest-frame at $z=3.3$) shows several faint sources ($> 23$ AB) embedded in the Ly$\alpha$ $3 \sigma$ detection limit emission (Figure \ref{fig:4_3}).
In particular, the peak of the Ly$\alpha$ emission lies very close ($0.6^{\prime \prime} \pm 0.3^{\prime \prime}$, or $4.6 \pm 2.3$~kpc at $z=3.3$), but not coincident, to the source labeled as ``A'' in Figure \ref{fig:4_3}. Although not shown here, similar spatial offset of source ``A'' is seen between $I$- and $g$-band images likely due to the strong contribution of the Ly$\alpha$ emission in the latter (see Figure \ref{fig:4_5}). VLA 1.4 GHz data also reveal faint emission on both sides of source  “A”. 
The symmetry of the radio components with respect to source ``A'' suggests that these could be radio jets or lobes associated with the galaxy ``A'', although a different configuration composed of two interacting AGNs (radio sources) with some leakage rest-frame UV light (source ``A'') cannot be ruled out with the available data \citep[e.g.,][]{ivison2007, rujopakarn2016, stach2019}. The scenario with two radio jets associated with ``A'' is nevertheless favoured by the detection of relatively bright extended emission in metal lines in the nebula (e.g., C~{\sc iv}) with perturbed kinematics within the radio structures (see Section \ref{kin}), similar to what is found in other HzRGs at similar redshifts \citep[e.g.,][]{villar2003, humphrey2006}. 

Source ``A'' is compact, but slightly resolved in the E-W direction in the Subaru image and in the higher spatial resolution {\it HST} image. 
A radial profile fitting of source ``A'' in the {\it HST} image yields a $\rm FWHM \simeq 0.4^{\prime \prime}$ or $\simeq 3$ kpc at $z = 3.3$, after correcting it for the intrinsic PSF measured using several stars in the field. The resolved spatial structure is also evident in the GTC spectrum (see left panel of Figure \ref{fig:4_4}).

\subsubsection{Emission lines and systemic redshift}\label{A}

\begin{figure*}[ht!]
\centering
$\begin{array}{rl}
    \includegraphics[width=0.98\textwidth]{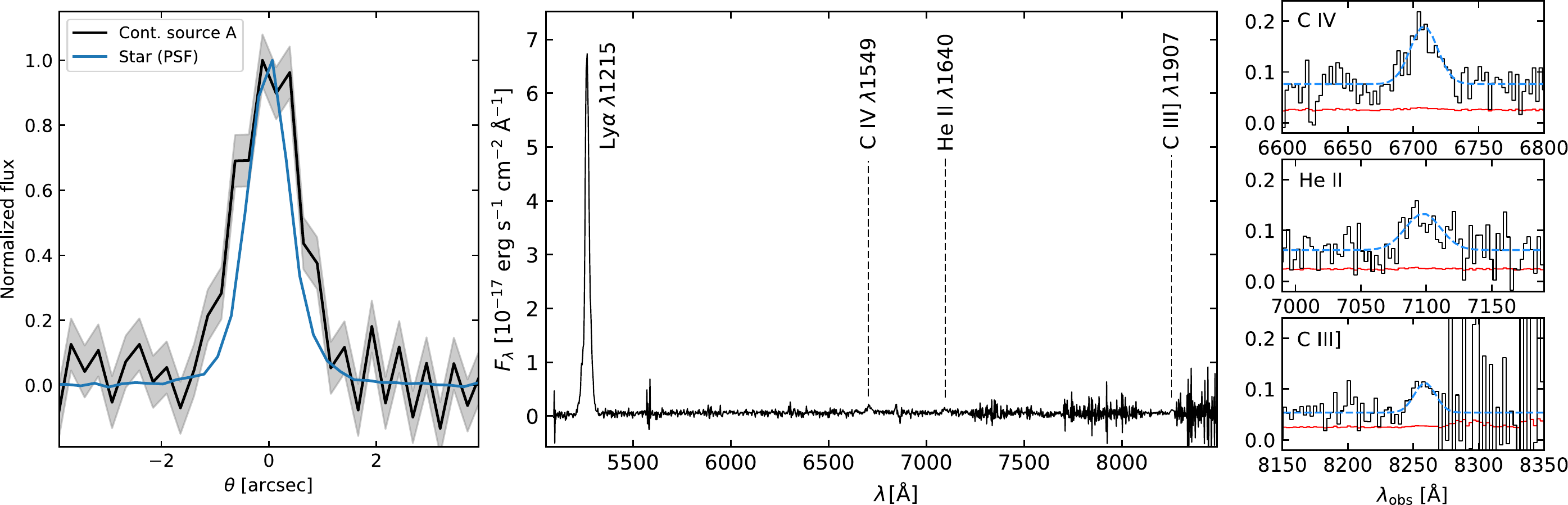}
\end{array}$
\caption{Left: Normalized 1D spatial profile (along the slit) of the UV continuum of source ``A'' (black solid line and 1$\sigma$ uncertainty in grey). The PSF profile measured using the reference star at $\sim 40^{\prime \prime}$ SE of HLock01-LAB is also shown in blue. The UV continuum of source ``A'' is compact, but slightly resolved in our spectrum, as well as in the Subaru image and in the high spatial resolution {\it HST} image. Middle: OSIRIS/GTC 1D spectrum of source ``A''. The spectrum is characterized by a strong emission in the Ly$\alpha$ line and a very faint continuum (barely detected). Emission in the C~{\sc iv}, He~{\sc ii}, and C~{\sc iii}] lines are also present, but these appear much fainter (marked with vertical dashed lines). 
N~{\sc v} 1238,1242\AA{ }is not detected within a $3\sigma$ limit of $2.2 \times 10^{-17}$ erg s$^{-1}$ cm$^{-2}$. 
Right panels show Gaussian fits (blue dashed lines) to the spectral profiles of C~{\sc iv}, He~{\sc ii}, and C~{\sc iii}] ($1 \sigma$ errors in red).
\label{fig:4_4}}
\end{figure*}

We extract the OSIRIS 1D spectrum of source ``A'' (from the $1.5^{\prime \prime}$-wide long slit) using a small aperture of $6$ pixels in the spatial direction ($\simeq 1.5^{\prime \prime}$, see Figure \ref{fig:4_3}). Figure \ref{fig:4_4} shows the 1D spectrum of source ``A''. It shows a strong and relatively broad Ly$\alpha$ emission with an observed flux $F_{\rm Ly \alpha}^{\rm obs} = (17.81 \pm 0.11) \times 10^{-16}$ erg s$^{-1}$ cm$^{-2}$ and a $\rm FWHM = 1400 \pm 150$ km s$^{-1}$, after accounting for the instrumental broadening ($\simeq 650$~km~s$^{-1}$), and a very faint rest-frame UV continuum.

We detect other emission lines, including C~{\sc iv}, He~{\sc ii}, and C~{\sc iii}], although more weakly than the Ly$\alpha$ line. By fitting a single Gaussian model to the line profiles, we measure fluxes of $F_{\rm C IV}^{\rm obs} = (2.8 \pm 0.4) \times 10^{-17}$, $F_{\rm He II}^{\rm obs} = (2.4 \pm 0.6) \times 10^{-17}$, and $F_{\rm C III]}^{\rm obs} = (0.9 \pm 0.3) \times 10^{-17}$ erg s$^{-1}$ cm$^{-2}$. 
Errors refer to $1\sigma$ uncertainties, estimated by perturbing independently 1000 times each spectra using the uncertainty of the flux of each spectral element.

Similar to Ly$\alpha$, C~{\sc iv} and He~{\sc ii} emission lines present relatively broad spectral profiles with $\rm FWHM = (960 \pm 160)$ and ($1200 \pm 300$) km s$^{-1}$, respectively (all values already corrected for the instrumental broadening). Despite the low significance of the detection ($\sim 3-7\sigma$), the measured FWHMs are too broad to be consistent with systemic rotation. On the other hand, the C~{\sc iii}] doublet emission appears spectrally unresolved, although we note that its red emission wing is highly affected by sky-subtracted residuals (see right panel of Figure \ref{fig:4_4}) making the measurement of its FWHM unreliable.
Nevertheless, given the line FWHMs ($<1500$ km~s$^{-1}$) and the weak continuum emission, source ``A'' can be classified as a type-II AGN  \citep[e.g.,][]{zakamska2003, alexandroff2013}. See Section \ref{diagnostic} for a more detailed study of the physical conditions of the ionized gas.

The redshift of source ``A'' is determined from the central wavelength of a Gaussian fit to the non-resonant He~{\sc ii} emission line. This yields the systemic redshift $z_{\rm A} = 3.326 \pm 0.002$. 
The spectrum does not show emission in N~{\sc v} 1238,1242\AA{ }(within a $3\sigma$ limit of $2.2 \times 10^{-17}$ erg s$^{-1}$ cm$^{-2}$) or N~{\sc iv} 1485\AA{ } and O~{\sc iii]} 1660,1666\AA{ } (within a $3\sigma$ limit of $1.5 \times 10^{-17}$ erg s$^{-1}$ cm$^{-2}$). Note however that, according to the AGN composite spectrum of \cite{hainline2011}, the aforementioned lines are expected to be much weaker than C~{\sc iv} (with C~{\sc iv}/N~{\sc v}~$ \simeq 3$, C~{\sc iv}/N~{\sc iv}~$ \simeq 8$, and C~{\sc iv}/O~{\sc iii]}~$ \simeq 15$).  Similarly, typical high-$z$ radio galaxies also show C~{\sc iv}/N~{\sc v}~$ \simeq 3$, C~{\sc iv}/N~{\sc iv}~$ \simeq 20$, and and C~{\sc iv}/O~{\sc iii]}~$ \simeq 7$ \citep[][]{humphrey2008}, and therefore consistent with the non-detection of these lines in the GTC spectrum (C~{\sc iv}/N~{\sc v}~$ > 1.3$, C~{\sc iv}/N~{\sc iv}~$ > 1.8$, and C~{\sc iv}/O~{\sc iii]}~$ > 1.8$, for a $3\sigma$ limit).

To infer the contribution of the Ly$\alpha$ emission associated with source ``A'' to the total luminosity of the nebula, we compare the normalized spatial profiles (extracted along the slit) of the total Ly$\alpha$ emission of the nebula and the UV continuum of source ``A'' (using the spectral region of $\simeq 1250 - 1500$~\AA). We find that the Ly$\alpha$ emission from source ``A'' contributes approximately $50$\% of the total emission of the nebula.

\subsubsection{Photometry and multi-wavelength analysis}\label{multi}

In this section we present the photometry of source ``A'' using the broad-band imaging data. These measurements are summarized in Table \ref{tab4_2}.
We use aperture photometry in the GTC $g$, Subaru $I$, {\it HST} WFC3 F110W, and WHT $K_{\rm s}$ bands. To do so, we measure the flux in an aperture with a diameter of $2.5 \times $ the PSF FWHM of each image. For $R$ band, we use the photometry from the corresponding CFHT/MEGACAM catalog downloaded from the Canadian Astronomy Data Centre (CADC\footnote{\url{http://www.cadc-ccda.hia-iha.nrc-cnrc.gc.ca/en/cfht/}}). 
For the {\it Spitzer} IRAC bands, we use the $3.8^{\prime \prime}$ aperture photometry provided by the {\it Spitzer} Enhanced Imaging Products (SEIP) catalog\footnote{\url{ http://irsa.ipac.caltech.edu/data/SPITZER/Enhanced/SEIP/}}, which includes catalogs from deep imaging data in the two first IRAC bands ($3.6$ and $4.5$ $\mu$m) from the {\it Spitzer} Extragalactic Representative Volume Survey \citep[SERVS:][]{mauduit2012}, and shallower IRAC ($5.8$ and $8.0$ $\mu$m) and MIPS ($24$ $\mu$m) imaging from the {\it Spitzer} Wide-Area InfraRed Extragalactic survey data \citep[SWIRE:][]{lonsdale2003}. 
However, source ``A'' is not detected in the IRAC $8.0$ $\mu$m and MIPS 24 $\mu$m bands, which correspond to upper limits of $38$ and $118$ $\mu$Jy ($5 \sigma$ confidence level), respectively.

\begin{table}[t!]
\begin{center}
\caption{Photometry of source ``A'' \label{tab4_2}}
\begin{tabular}{l c c}
\hline \hline
\smallskip
\smallskip
Telescope/Band  &  $\lambda_{\rm obs}$  &  Flux \\
      &   ($\mu$m)      &  ($\mu$Jy) \\
\hline 
GTC/$g$-band  & 0.48 & $2.1 \pm 0.2$ \\
CFHT/$r$-band  & 0.62 & $1.6 \pm 0.1$ \\
Subaru/$I$-band   & 0.77 & $1.8 \pm 0.1$ \\
{\it HST}/F110W & 1.12 & $3.3 \pm 0.4$ \\
WHT/$K_{\rm s}$ & 2.16 & $14.2 \pm 0.9$ \\
{\it Spitzer}/IRAC I1 & 3.6 & $28.0 \pm 0.3$ \\
{\it Spitzer}/IRAC I2 & 4.5 & $32.6 \pm 0.6$ \\
{\it Spitzer}/IRAC I3 & 5.8 & $37 \pm 9$ \\
{\it Spitzer}/IRAC I4 & 8.0 & $\leq 38$ ($5 \sigma$)\\
{\it Spitzer}/MIPS M1 & 24.0 & $\leq 118$  ($5 \sigma$) \\
SMA     & 880  & $\leq 4100$  ($5 \sigma$) \\
CARMA   & 3300 & $\leq 830$  ($5 \sigma$) \\
VLA 1.4~GHz     & 214000 & $280 \pm 40$ \\
\hline 
\end{tabular}
\\
\end{center}
\textsc{      \bf{}} \\
 \\
\end{table}

\begin{figure}[ht!]
\centering
$\begin{array}{rl}
    \includegraphics[width=0.45\textwidth]{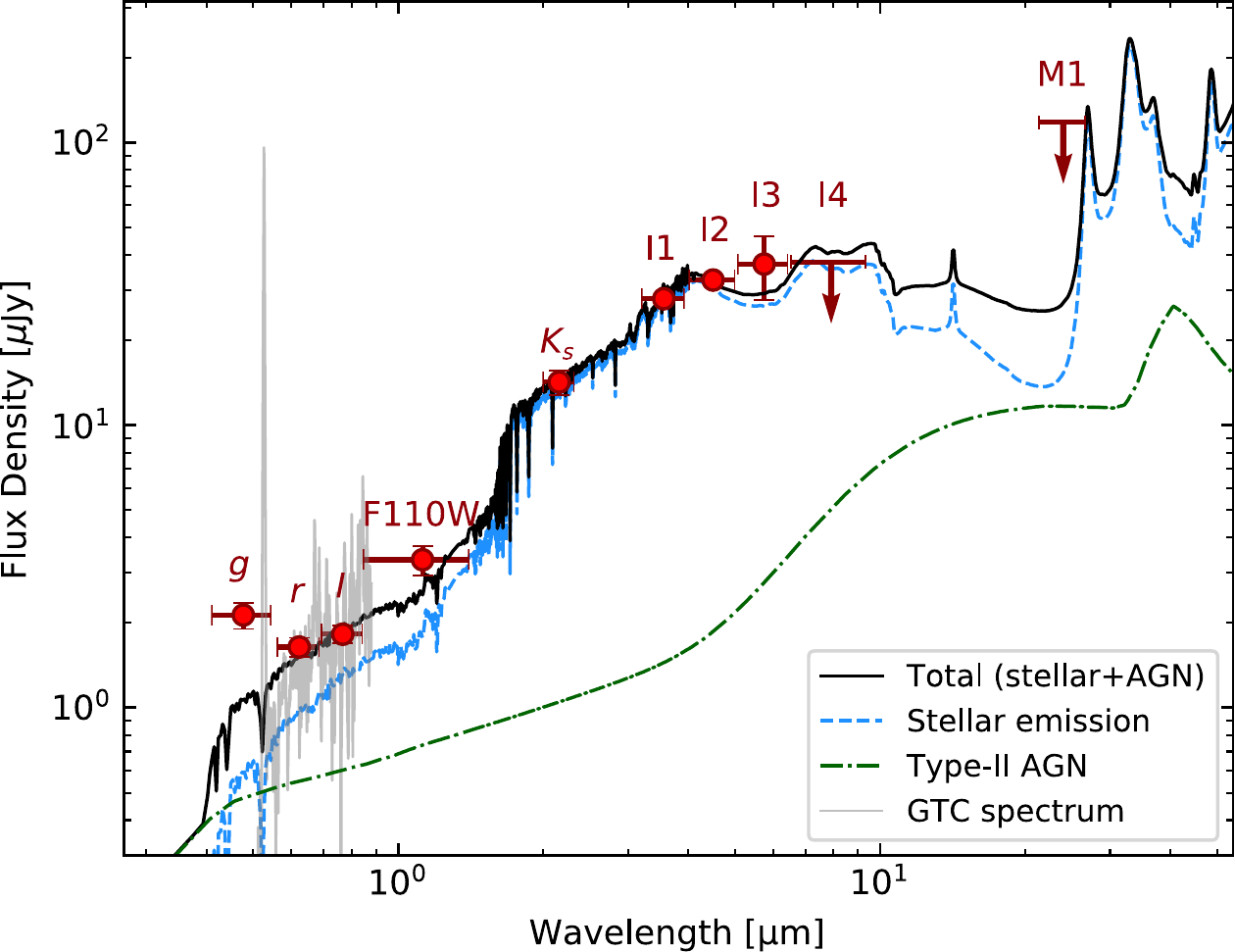}
\end{array}$
\caption{Best-fit model of the spectral energy distribution of source ``A'' using {\sc SED3fit} \citep{berta2013}. Black line is the total fitted emission, whereas blue and green lines represents the contribution of stellar and AGN emission, respectively. 
The fit uses photometry from CFHT $r$ to IRAC $5.8$ $\mu$m, and the non detection in the {\it Spitzer} IRAC $8.0$ $\mu$m and MIPS $24$ $\mu$m bands (in red). We also show the GTC spectrum of source ``A'' (in grey, smoothed for visual purpose). Note the flux excess in the $g$-band (not included in the fit) due to the strong Ly$\alpha$ emission. 
\label{fig:4_5}}
\end{figure}

To investigate the contribution of an AGN in the SED of source ``A'', we use the multi-component SED fitting tool {\sc Sed3fit} \citep{berta2013}.
This code is based in the \cite{dacunha2008} {\sc Magphys} code and employs the combination of stellar emission, dust emission from star-forming regions, and emission from a type-I/II AGN (AGN torus libraries from an updated version of the \citealt{fritz2006} models by \citealt{feltre2012}; see also: \citealt{gruppioni2016, delvecchio2017, delvecchio2018}). 
The fit uses optical $R$ and $I$, near-IR F110W and $K_{\rm s}$, and mid-IR {\it Spitzer} $3.6$, $4.5$, and $5.8$ $\mu$m flux measurements, along with the detection limits of IRAC~$8$~$\mu$m, MIPS~$24$~$\mu$m, and SMA~880$\mu$m images. 
We exclude the photometry of the GTC $g$ band in the fit, given the large contribution of the Ly$\alpha$ emission (see Figure \ref{fig:4_5}). At $z=3.326$, H$\beta$~4862\AA{ }and [O~{\sc iii]}~4960,5008\AA{ }are redshifted to the $K_{\rm s}$ band. However, in this case, we do use the $K_{\rm s}$ flux measurement in the fit, as we do not expect a large contribution of these rest-frame optical emission lines in the photometry ($\simeq 0.04$ mag, assuming typical line ratios of H$\beta$/C~{\sc iv}~$\simeq 0.5$ and [O~{\sc iii]}/C~{\sc iv}~$\simeq 4.4$ found in other HzRGs \cite{humphrey2008}).

Stellar population synthesis models of \cite{bruzual2003}, the \cite{chabrier2003} initial mass function, and an exponentially declining star formation history (i.e., $\propto e^{-t/\tau}$) are assumed.

The best-fit model ($\chi^{2} = 2.3$), shown in Figure \ref{fig:4_5}, gives a small AGN contribution ($\sim 10 \%$) to the total light emission of source ``A'' between 8 and 1000~$\mu m$ (rest-frame). We derive a stellar mass of source ``A'' as $\rm log ( M_{*}/M_{\odot}) = 11.37 \pm 0.09$ with age $\rm log( \rm age_{M}/ yr^{-1}) =  8.6\pm0.2$, metallicity $Z/Z_{\odot} = 1.09$, and dust attenuation $A_{\rm V} = 0.55 \pm 0.25$~mag. Similar to the host galaxies of powerful radio sources, source ``A'' is  extremely massive \citep[e.g.,][]{rocca2004, seymour2007}. The star formation rate is found to be $\rm SFR = 50_{-10}^{+170}$ $M_{\odot}$ yr$^{-1}$. 

On the other hand, the properties of the AGN component are much less constrained due to the lack of deep mid-IR data \citep[see][]{berta2013}. Nevertheless, the best-fit model gives an IR luminosity (measured from the rest-frame 8 to 1000 $\mu$m) for the AGN component $L_{\rm IR}^{\rm AGN} \simeq 3 \times 10^{10}$~$\rm L_{\odot}$.

\subsubsection{Far-IR and radio emission}\label{radio}

Deep VLA 1.4 GHz data with $\simeq 1.1^{\prime \prime}$ resolution \citep[see:][]{wardlow2013, marques2018} show faint emission on both sides of source ``A'' with similar intensities, suggestive of radio jets or lobes (red contours in Figures \ref{fig:4_2} and \ref{fig:4_3}). 
The radio components show flux densities $S_{\rm West} = 0.151 \pm 0.03$~mJy and $S_{\rm East} = 0.116 \pm 0.03$~mJy for the West and East counterparts, respectively. At the position of HLock01-LAB, we do not detect any emission in the SMA 880 $\mu$m image at a significance level $5 \sigma = 4.1$ mJy. Following \cite{magnelli2015}, we use the empirical far-IR/radio correlation to study the radio excess in HLock01-LAB. The $q_{\rm FIR}$ parameter is defined as:

\begin{equation}\label{eq1}
q_{\rm FIR} = \log  \left( \frac{L_{\rm FIR} [\rm W]}{3.75 \times 10^{12}} \right)  - \log(L_{\rm 1.4 GHz} [\rm W~Hz^{-1}]),
\end{equation} 

\noindent
where $L_{\rm FIR}$ is the integrated luminosity from the rest-frame 42 to 122 $\mu$m, and $L_{\rm 1.4 GHz}$ is the rest-frame 1.4~GHz radio luminosity \citep[we assume a radial spectral index $\alpha = -1$, where $S_{\nu} \propto \nu^{\alpha}$;][]{smolcic2017}. 
By re-scaling the average far-IR spectral energy distribution of ALESS galaxies \citep{dacunha2015} to the SMA 880 $\mu$m $5 \sigma$ limit, we find a FIR luminosity $L_{\rm FIR} < 2.3 \times 10^{12}$ $L_{\odot}$.\footnote{This corresponds to a $L_{\rm FIR} < 4.0 \times 10^{12}$~$L_{\odot}$ integrated from 8-1000$~\mu$m in the rest-frame, implying an upper limit of $\rm SFR < 690$~M$_{\odot}$~yr$^{-1}$ following \cite{kennicutt1998}.}
Using equation \ref{eq1} this yields to a $q_{\rm FIR} < 0.92$, much lower than the mean value and scatter of $q_{\rm FIR} = 2.3 \pm 0.7 $ typically found in other star-forming galaxies \citep[e.g.,][]{yun2001, ivison2010}. Thus, the radio exccess in HLock01-LAB clearly indicates the presence of an AGN.

\subsection{Kinematics of the ionized gas}\label{kin}

\begin{figure}[ht!]
\centering
$\begin{array}{rl}
    \includegraphics[width=0.40\textwidth]{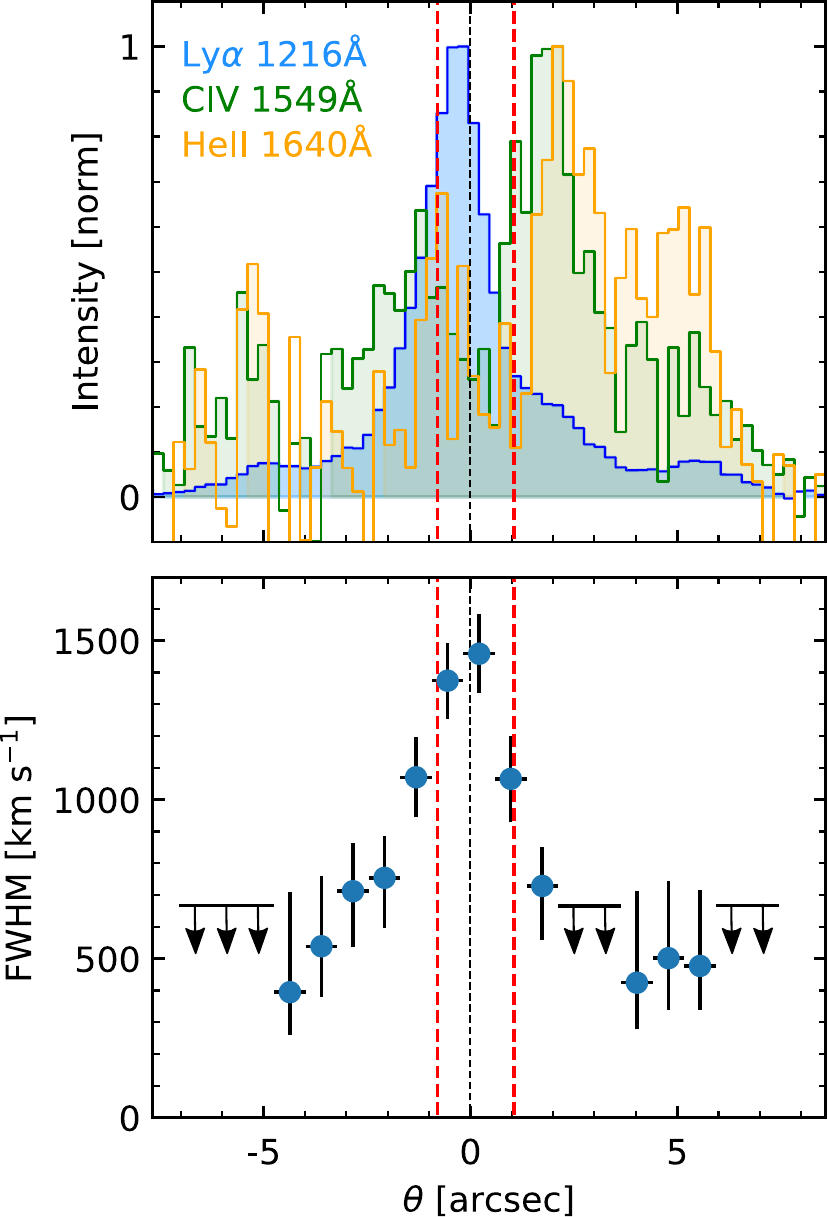}
\end{array}$
\caption{Top: Spatial distribution of the normalized flux along the slit of Ly$\alpha$ (blue), C~{\sc iv} (green), and He~{\sc ii} (orange). Negative and positive values of $\theta$ correspond to $\rm PA=110^{\circ}$ and $290^{\circ}$ (measured north to east), respectively. Continua have been subtracted.
Bottom: Ly$\alpha$ spectral FWHM (already corrected from the instrumental broadening).  
Arrows represent the FWHM upper limits ($\simeq 650$~km~s$^{-1}$). The spatial zero marks the location of source ``A'', whereas the approximate positions of VLA radio components are marked with red dashed lines. 
\label{fig:4_6}}
\end{figure}

Despite the low spectral resolution of the R1000R grism, the visual inspection of the 2D spectra reveals variations on the kinematics of the gas along, and well beyond the radio structures. 
Figure \ref{fig:4_6} shows the spatial distribution of the normalized flux along the slit of Ly$\alpha$, C~{\sc iv}, and He~{\sc ii} (the continuum has been subtracted in all lines). It is worth noting that Ly$\alpha$ emission drops sharply where both C~{\sc iv} and He~{\sc ii} reach their maximum, at approximately $2^{\prime \prime}$ NW of source ``A'' (knot ``B''). 

In order to study the kinematics of the Ly$\alpha$ emission,\footnote{The C~{\sc iv} and He~{\sc ii} extended emission presents low S/N, thus we do not investigate their kinematics.} one dimensional spectra were extracted from different apertures along the slit with sizes of 3 pixels each (corresponding to $\sim 0.76^{\prime \prime}$) and the Ly$\alpha$ lines were fitted with Gaussian profiles.
FWHM and velocities of Ly$\alpha$ line are also shown in Figure \ref{fig:4_6}. Values of FWHM have been corrected for the instrumental broadening ($\simeq 650$ km s$^{-1}$). 

Perturbed kinematics, i.e. $\rm FWHM \gtrapprox 1000$ km s$^{-1}$, are detected preferentially in the inner region between the two radio components (red dashed lines in Figure \ref{fig:4_6}), and reaches its maximum $\rm FWHM \simeq 1400$ km s$^{-1}$ around source ``A'' (black dashed line). 
We are confident that these results are solid, and not dependent on slit effects (see Section \ref{sec:2.2}). On one hand, the giant nebula probably fills the slit (see Figure \ref{fig:4_3}), thus, the kinematics of the large scale gas beyond the radio structures is not affected by slit effects. The Ly$\alpha$ emission associated with galaxy A is more compact, but slit effects would not affect our conclusions, since the dominant source of line broadening is kinematic, rather than instrumental. On the other hand, accounting for slit effects would result, if anything,  on slightly broader lines within the radio structures \citep[][]{villar2000}. 
In addition to the perturbed kinematics, the Ly$\alpha$ line presents also high surface brightness within the radio components ($\sim 10^{-16}$~erg~s$^{-1}$~cm$^{-2}$~arcsec$^{-2}$). Both the perturbed kinematics and the high surface brightness highly suggest a strong interaction between the radio jets or lobes and the surrounding gas, as seen in other Ly$\alpha$ nebula around powerful HzRGs \citep[e.g.,][]{villar1999, bicknell2000, humphrey2006}.

In the outer regions of the nebula the Ly$\alpha$ line shows a narrower profile with FWHM~$< 650$~km~s$^{-1}$ (down to $\simeq 400$~km~s$^{-1}$ in some regions) and much lower surface brightness (few times $10^{-18}$~erg~s$^{-1}$~cm$^{-2}$~arcsec$^{-2}$) than in the central region of the nebula. Such properties resemble those found in quiescent halos around HzRGs and other radio-quiet sources, that are thought to have collapsed in early phases of galaxy formation \citep{villar2003}. The presence of heavy elements (C~{\sc iv}) in the outer regions of the nebula also indicates that feedback, associated either with star-formation or AGN outflows, may have enriched the halo at least 50~kpc from the nuclear region (source ``A'').

Note, however, that Ly$\alpha$ profile is sensitive to absorption by neutral hydrogen and the kinematic analysis based on this line should be treated with care. 
The spectral resolution of our data is insufficient to further investigate the presence of strong H~{\sc i} absorbers over the full spatial extent of the Ly$\alpha$ nebula, although the higher spectral resolution data used in the analysis of HLock01 in \cite{marques2018} show that the blue wing of the Ly$\alpha$ emission $\simeq 2^{\prime \prime}$ SW of source ``A'' is heavily absorbed (see Figure \ref{fig:4_1}).\footnote{The red wing of the Ly$\alpha$ emission also shows an absorption line at $\simeq 5281$\AA, likely associated with C~{\sc ii} 1334\AA{ }of the $z=2.957$ SMG HLock01 \citep{marques2018}.} 
Nevertheless, this effect is not supposed to change our conclusions regarding the striking differences between the gas kinematics within (turbulent) and outside (more quiescent) the radio structures, since the extended halo is not expected to be so severely affected by absorption as the central region.

\subsection{Line diagnostics in the nebula}\label{diagnostic}

\subsubsection{AGN versus star formation}

To gain insight into the physical conditions of the ionized gas traced by C~{\sc iv}, He~{\sc ii}, and C~{\sc iii}] nebular emission, we use rest-frame UV emission-line diagnostics to identify the source of photoionization. Following \cite{nakajima2017}, the line ratios of C~{\sc iv}/C~{\sc iii}] (C4C3) and (C~{\sc iii}]$+$C~{\sc iv})/He~{\sc ii} (C34) can be used to distinguish star-forming galaxies from AGNs \citep[see also][]{feltre2016}. These photoionization models were constructed using a grid with ionizing parameter (log~$U$) ranging from $-3.5$ to $-0.5$, along with different gas properties of metallicity ($Z/Z_{\odot}$ ranging from $10^{-4}$ to $5.0$) and density ($n \sim 10 - 10^{5}$~cm$^{-3}$), thus they are also valid for distant non-galaxy regions, where the gas is expected to be much more diluted.

In addition to the spectrum of source ``A'', we extract 1D spectra from other regions of the nebula. These include the region encompassing the radio emission (core), the NW and SE extended regions of the nebula (NW and SE halos, respectively), and additional knots ``B'' and ``C'' where C~{\sc iv} and He~{\sc ii} fluxes are relatively large. The location and size of these apertures are shown in Figure \ref{fig:4_3}, as well as in Table \ref{ratios_tab}.

\begin{table*}[t!]
\begin{center}
%\tabletypesize{\scriptsize}
\caption{Ly$\alpha$ fluxes and emission-line ratios of HLock01-LAB.\label{ratios_tab}}
\begin{tabular}{l c c c c c c c c}
\hline \hline
\smallskip
\smallskip
Region  &  Aperture &  $F$ (Ly$\alpha$) & FWHM (Ly$\alpha$) & Ly$\alpha$/N~{\sc v} & Ly$\alpha$/C~{\sc iv} & Ly$\alpha$/He~{\sc ii} &  Ly$\alpha$/C~{\sc iii}] & C~{\sc iv}/ He~{\sc ii}\\
   & ($^{\prime \prime}$)  & ($10^{-17}$ erg~s$^{-1}$~cm$^{-2}$)   &  (km s$^{-1}$) &    &   &  & &  \\
\hline 
Source ``A'' & $1.5$ & $178.1   \pm 1.1$ & $1400 \pm 150$ & $>80$  & $64 \pm 10$  & $75 \pm 20$  &  $200 \pm 80$ & $1.2 \pm 0.4$ \\
Core         & $2.5$ & $238.2 \pm 1.3$ & $1300 \pm 150$ & $>87$ & $40 \pm 4$   & $82 \pm 33$  &  $174 \pm 110$ & $2.1 \pm 0.9$ \\
knot ``B''   & $1.0$ & $22.6  \pm 0.7$ & $750  \pm 200$ & $>10$ & $5.7 \pm 0.6$& $13 \pm  3$  &  $> 20$  & $2.3 \pm 0.6$ \\
knot ``C''   & $1.0$ & $5.3   \pm 0.4$ & $< 650$        & $>3$ & $4.0 \pm 0.9$& $8  \pm  5$  &  $> 5$  & $2.1\pm1.4$\\
NW halo      & $5.1$ & $23    \pm 1.3$ & $< 650$        & $>5$ & $6   \pm 2 $ & $8  \pm  3$  &  $>10$  & $1.4 \pm 0.7$\\
SE halo      & $2.5$ & $35.6  \pm 0.9$ & $650 \pm 200$  & $>12$  & $10  \pm 2$  & $>15$        &  $>15$ & $>1.5$\\
Total        & $16.3$ & $385  \pm 5$ & $1000 \pm 200$ & $>31$ & $16  \pm 1$  & $43 \pm 13$  &  $> 25$  & $2.6 \pm 0.8$\\
\hline 
%Mean         &           &          &    2.9545$\pm$0.0004   \\   
%\hline 
\end{tabular}
\\
\end{center}
\textsc{      \bf{Notes:}} Lower limits refer to $3\sigma$ assuming a $\rm FWHM = 1000$~km~s$^{-1}$. \\
 \\
\end{table*}

\begin{figure}[ht!]
\centering
$\begin{array}{rl}
    \includegraphics[width=0.46\textwidth]{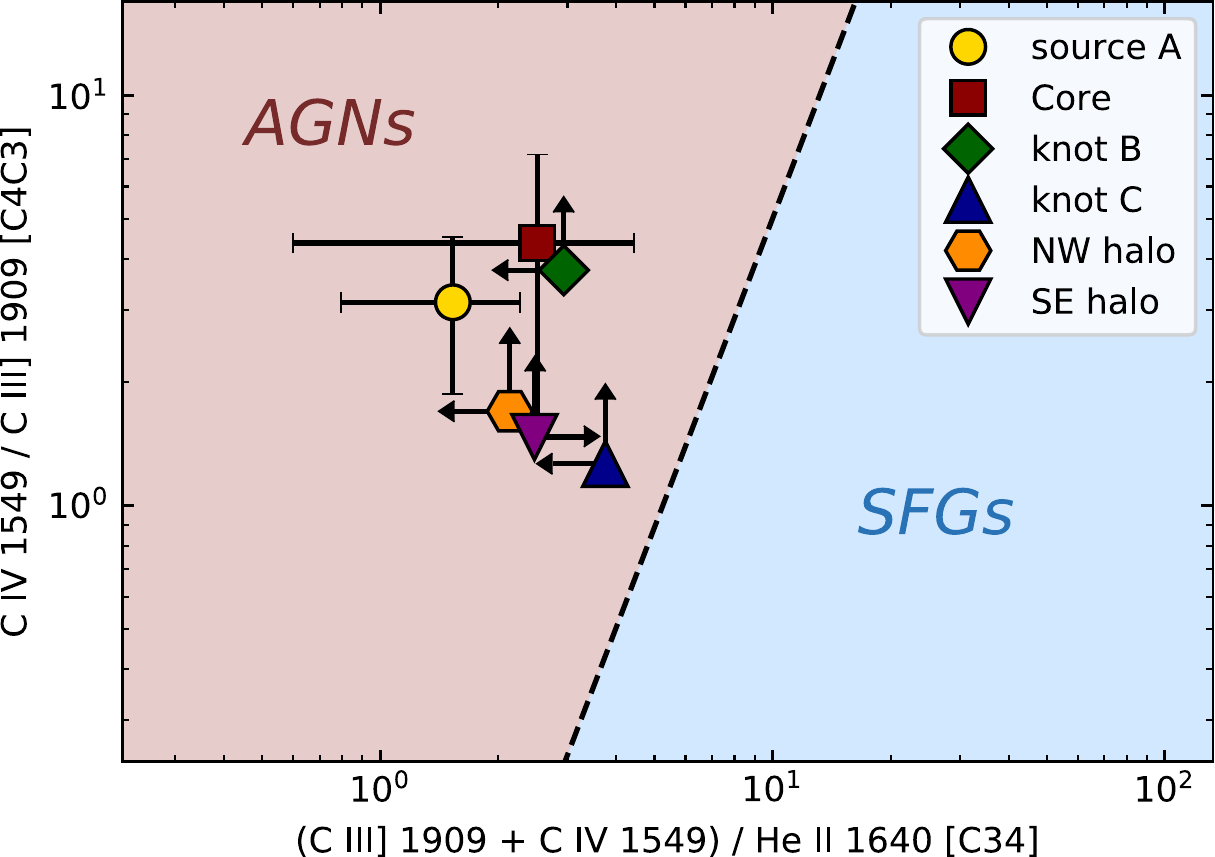}
\end{array}$
\caption{Positions of several components of HLock01-LAB in the diagram of C4C3 vs. C34 proposed by \cite{nakajima2017}. The line ratios of C~{\sc iv}/C~{\sc iii}] (C4C3) and (C~{\sc iii}]$+$C~{\sc iv})/He~{\sc ii} (C34) clearly show that the gas is excited by an AGN and not by star formation.
\label{fig:naka}}
\end{figure}

From the GTC spectra, we find line ratios of $\rm C34 = (1.5 \pm 0.7)$, ($2.5 \pm 1.9$), $<2.9$, $<3.7$, $<2.1$, and $>2.4$, and $ \rm C4C3 = (3.1 \pm 1.3$), ($4.4 \pm 2.8$), $>3.8$, $>1.3$, $>1.7$, and $>1.5$ for source ``A'', core, knots ``B'' and ``C'', and NW and SE halos, respectively. As shown Figure \ref{fig:naka}, these ratios suggest that the ionized gas in the nebula is powered by an AGN, likely source ``A'', rather than star formation. 

In addition, we also use the rest-frame equivalent widths ($EW_{0}$) of C~{\sc iv} and C~{\sc iii]} combined with the line ratios of C~{\sc iv}/He~{\sc ii} and C~{\sc iii]}/He~{\sc ii} to disentangle AGN from star-formation activity as proposed by \cite{nakajima2018} (see also \citealt{hirschmann2019}). From the spectrum of source ``A'' (where the continuum emission is detected) we measure $EW_{0}^{\rm CIV} = 20 \pm 9$ and $EW_{0}^{\rm CIII]} = 11 \pm 8$, and C~{\sc iv}/He~{\sc ii}~$=1.2 \pm 0.4$ and C~{\sc iii]}/He~{\sc ii}~$=0.4 \pm 0.2$, indicating that the gas is excited by an AGN.

On the excitation of Ly$\alpha$ in the outer regions of HLock01-LAB, the detection of both C~{\sc iv} and He~{\sc ii} over similar extension as Ly$\alpha$ makes resonant scattering of Ly$\alpha$ or cooling radiation from pristine gas unlikely scenarios \citep[see:][]{arrigoni2015b}.

\subsubsection{Extreme Ly$\alpha$/C{\sc iv} and Ly$\alpha$/He{\sc ii} emission line ratios}\label{line_ratios}

The inner region (i.e. core and source ``A'') of HLock01-LAB, encompassing the AGN and the radio structures, shows extremely large emission line ratios of Ly$\alpha$/C{\sc iv} and Ly$\alpha$/He{\sc ii}, up to $64 \pm 10$ and $82 \pm 33$, respectively (see Table \ref{ratios_tab}). Quasars, radio galaxies and type-II AGNs at similar redshifts show significantly lower values (Figure \ref{fig:4_7}). Ly$\alpha$ is enhanced both relative to other emission lines and in absolute terms. For comparison, HzRGs, which host more powerful AGNs, have similar Ly$\alpha$ luminosities. 

AGN photoionization models covering a broad range of gas densities, metallicities and ionization parameters  predict Ly$\alpha$/He{\sc ii}~$\la$~30 \citep{villar2007}, unless a combination of very low metallicities ($Z/Z_{\odot} \la$~0.1) and low ionization parameter values (log $U<10^{-4}$) are considered \citep[see also:][]{humphrey2019}. These models would result on C~{\sc iv}/He~{\sc ii}~$\ll$~1, which is inconsistent with the measured values ($1.2-2.1$, see Table \ref{ratios_tab}). The density in the central region is probably significantly higher than that in the very extended nebula, where the gas is expected to by much more diluted. Indeed, a broad range of densities is possible in the Narrow Line Region of AGNs ($ n \sim 100 - 10^{6}$~cm$^{-3}$, e.g., \citealt{osterbrock1989}). However, high densities cannot explain the strong Ly$\alpha$ emission relative to other lines. Even in the most extreme case ($n = 10^{6}$~cm$^{-3}$), models predict Ly$\alpha$/He{\sc ii}~$\sim 30$ (see Figure 3 in \citealt{villar2007}). On the other hand, the same models predict Ly$\alpha$/C~{\sc iv}~$\la$~10 and C~{\sc iv}/He{\sc ii}~$>3$, which are in contradiction with the observed ratios in the central region of HLock01-LAB  (see Table \ref{ratios_tab}).

This suggests that changing the gas properties such as density or metallicity cannot explain the Ly$\alpha$ enhancement. Instead, excitation mechanisms rather than pure AGN photoionization have to be taken into account.

The addition of stellar photoinization to the effects of the AGN would result on a softer ionizing continuum, that would enhance the Ly$\alpha$ luminosity and its ratios relative to C~{\sc iv} and He~{\sc ii} \citep{villar2007}. This process, however, cannot explain HLock01-LAB. Using \cite{kennicutt1998} calibration, the inferred $\rm SFR \sim 50$~M$_{\odot}$~yr$^{-1}$ (Section \ref{multi}) would result on  $L_{\rm Ly\alpha} \sim 5.5 \times 10^{43}$~erg~s$^{-1}$ \citep[assuming case B Ly$\alpha$/H$\alpha$=8.7,][]{valls1993}, provided that Ly$\alpha$ is not quenched by dust. Therefore, even in the most favourable conditions, star formation could account for less than 10\% of the total line luminosity.

Ly$\alpha$ collisional excitation is a more promising possibility. For this to happen, the electrons in the ground level of hydrogen must be excited by electrons with $kT\ga 10.2$~eV, and electron temperatures $T \ga 1.2 \times 10^{5}$~$K$ are thus necessary. 
The effect of collisional excitations upwards from the $n=1$ levels of H can have a dominant effect in astrophysical shocks \citep[][]{raga2015}. As explained by these authors, immediately after the shock, one has a high temperature region (of $\sim 10^{5}$~$K$ for a 100~km~s$^{-1}$ shock) in which H can be partially neutral, though rapidly becoming collisionally ionized. In this region, H $1 \to n$ collisional excitations dominate over the recombinations to the excited levels \citep[see also][]{raymond1979}. 

The fact that the large Ly$\alpha$/C{\sc iv} and Ly$\alpha$/He{\sc ii} ratios are seen only in the region encompassing the radio structures, where perturbed kinematics are also found (see Section \ref{kin}), strongly supports that jet-induced shocks are contributing to the enhancement of Ly$\alpha$.
To investigate this, we use the shock models presented already by \cite{arrigoni2015b}. These models are based on libraries of radiative shock models using the code MAPPINGS~III \citep{allen2008},\footnote{\url{http://cdsweb.u-strasbg.fr/~allen/mappings_page1.html}} and assume solar metallicity gas, a magnetic parameter $B/n^{1/2} = 3.23 \mu$G, and a grid with gas densities from 0.01 to 100~cm$^{-3}$ and shock velocities from 100 to 1000~km~s$^{-1}$. 
The ionizing flux strongly depends on the shock velocity ($F_{\rm UV} \propto v_{\rm s}^{3}$), yielding to gas temperatures as high as $10^{6}$~$K$ \citep[see][]{allen2008}. 
Values of Ly$\alpha$/C{\sc iv}$\sim 60$ and Ly$\alpha$/He{\sc ii}$\sim 80$ can be reached in models with shock velocities of $\sim 600-800$~km~s$^{-1}$ and densities $\sim 100$~cm$^{-3}$ \citep[see Figure 13 of][]{arrigoni2015b}. Shock models can also explain the high C~{\sc iv}/C~{\sc iii]}~$>3$ observed within the radio structures.

Therefore, shock-heating induced by the radio jets is a natural explanation for the enhanced Ly$\alpha$ emission in the inner region of HLock01-LAB. Deep integral field spectroscopy would be very valuable to investigate this scenario in more depth, by mapping the kinematic, ionization and morphological properties of HLock01-LAB in two spatial dimensions. This study will be presented in a future paper.

\section{Comparison with other Ly$\alpha$ nebulae}\label{sec:4}

In this section we compare the properties of HLock01-LAB with those from other Ly$\alpha$ nebulae.
A summary of the properties of HLock01-LAB is presented in Table \ref{tab4_3}.

\begin{table}[t!]
\begin{center}
%\tabletypesize{\scriptsize}
\caption{Properties of HLock01-LAB. \label{tab4_3}}
\begin{tabular}{l  c c c}
\hline \hline
\smallskip
\smallskip
  &  Value  & Uncertainty  & Unit \\
\hline 
R.A.$^{a)}$ & 10:57:49.74 & $0.2^{\prime \prime}$  & J2000    \\
Dec.$^{a)}$  & +57:30:15.0  & $0.2^{\prime \prime}$ & J2000 \\
$z$  & $3.326$ & $0.002$ & --- \\
Extension$^{b)}$  & $\sim 110$ & --- & kpc \\
$L_{\rm Ly\alpha}^{b)}$ & $6.4 \times 10^{44}$ & $0.1 \times 10^{44}$ & erg s$^{-1}$ \\ 
$L_{\rm IR}^{a, b)}$ & $< 2.3 \times 10^{12} $ & ---  & $L_{\odot}$ \\
$L_{\rm 1.4 GHz}^{b)}$ & $2.8 \times 10^{25} $ & $0.4 \times 10^{25} $  & W~Hz$^{-1}$ \\
$M_{*}^{a, b)}$ & $2.3 \times 10^{11} $ & $0.7 \times 10^{11} $ & $\rm M_{\odot}$ \\
\hline 
\end{tabular}
\\
\end{center}
\textsc{      \bf{Notes:}} $a)$ refers to source ``A''; $b)$ uncorrected for lensing magnification ($\mu < 1.5$). \\
 \\
\end{table}

\begin{figure}[ht!]
\centering
$\begin{array}{rl}
    \includegraphics[width=0.41\textwidth]{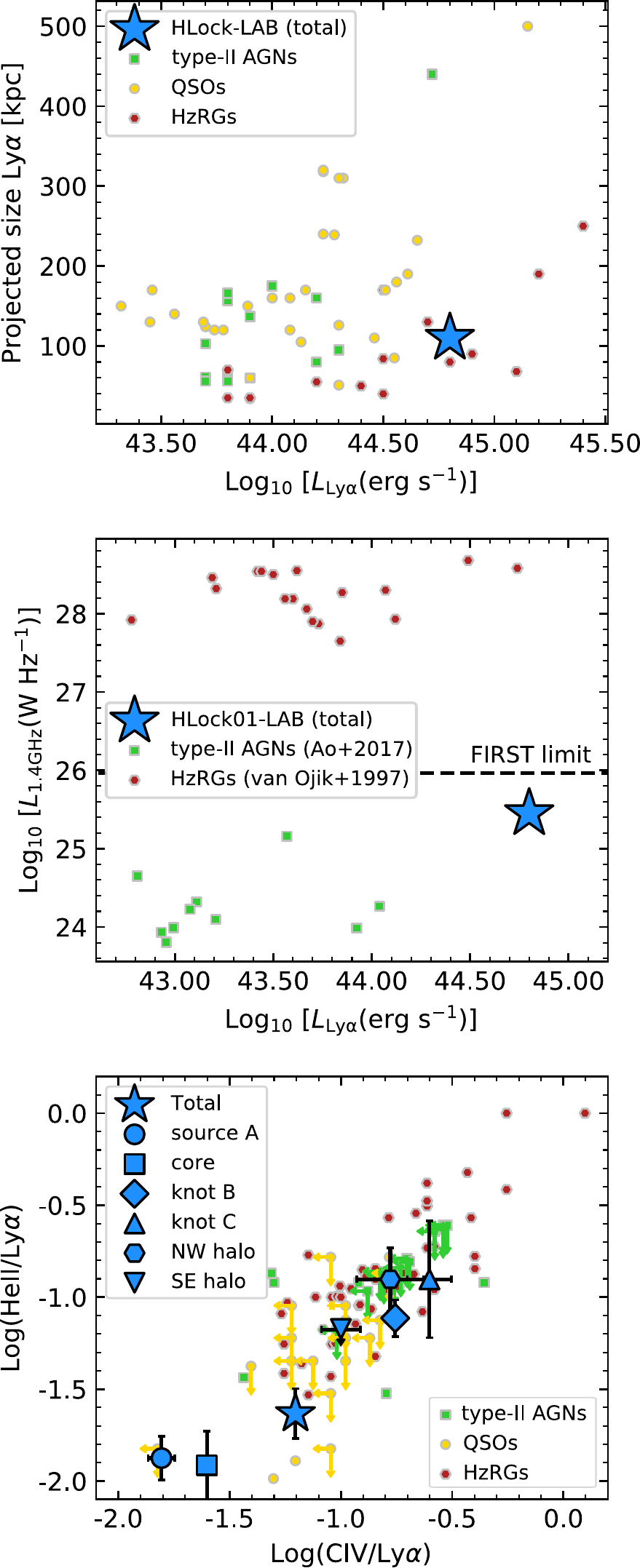}
\end{array}$
\caption{Up: Ly$\alpha$ luminosity and maximum projected extension of HLock01-LAB compared to other Ly$\alpha$ nebula associated with type-II AGNs, QSOs, and HzRGs (see references in the text). Middle: relation between the total radio power and Ly$\alpha$ luminosity of nebulae associated with HzRGs \citep{vanojik1997} and type-II AGNs \citep{ao2017}. The horizontal dashed line marks the approximate FIRST radio luminosity limit at $z\sim 3$ ($S_{1.4 \rm GHz} \sim 0.9$ mJy). Down: line ratios of Ly$\alpha$, C~{\sc iv}, and He~{\sc ii} of several regions of HLock01-LAB. Other Ly$\alpha$ nebula associated with type-II AGNs \citep{dey2005, prescott2009, prescott2013, arrigoni2015b, cai2017}, QSOs \citep{borisova2016, marino2019}, and HzRGs \citep{villar2007} are also shown. Detection limits of $2\sigma$ found in the literature have been converted to $3\sigma$. 
\label{fig:4_7}}
\end{figure}

HLock01-LAB has a total Ly$\alpha$ luminosity $L_{\rm Ly\alpha} = (6.4 \pm 0.1) \times 10^{44}$ erg s$^{-1}$ extended over $\simeq 110$ kpc. Even considering a possible lensing magnification ($\mu \sim 1.5$) from the group of $z=0.64$ galaxies at $\sim 15^{\prime \prime}$~SW, HLock01-LAB is one of the most luminous nebulae know at high redshift.
In Figure \ref{fig:4_7} we compare the maximum projected size of Ly$\alpha$ emission and the total luminosity of HLock01-LAB with a compilation of other giant Ly$\alpha$ nebulae associated with QSOs \citep[][]{cantalupo2014, hennawi2015, borisova2016, cai2018, arrigoni2018}, HzRGs \citep{vanojik1997, reuland2003, villar2003, villar2007a, venemans2007}, and type-II AGNs \citep[][]{overzier2013, ao2017, cai2017}.
Ly$\alpha$ halos around powerful radio-galaxies show statistically larger Ly$\alpha$ luminosity and broader kinematics with respect to other radio-quiet systems \citep[e.g.,][]{heckman1991b, miley2006}.

As discussed already in Section \ref{sec:3}, HLock01-LAB shares several of its properties with those found in other HzRGs at similar redshifts.
The Ly$\alpha$ morphology probed by the medium-band SHARDS image is apparently aligned with the radio axis, and the gas within the radio structures shows higher surface brightness and very perturbed kinematics. Such properties have been also found in other powerful HzRGs \citep[e.g.,][]{mccarthy1987, mccarthy1995, villar2003, villar2007a, morais2017}, and have been interpreted as further evidence of the jet-gas interaction that distorts the morphological and kinematic properties of the surrounding gas \citep[e.g.,][]{villar1999, bicknell2000, humphrey2006}. However, despite the general similarities between HLock01-LAB and other giant and luminous Ly$\alpha$ nebulae around powerful HzRGs, there are striking differences that should be discussed. 

First of all, the radio emission seen in HLock01-LAB is much weaker (total flux density $S_{1.4 \rm GHz} = 0.27 \pm 0.04$ mJy) than those found in other HzRGs at similar redshifts, showing typically very strong radio emission with flux densities up to hundreds of mJy or more  \citep[e.g.,][]{roettgering1994, vanojik1997, debreuck2000, debreuck2004}. Figure \ref{fig:4_7} shows the relation between the total radio power and Ly$\alpha$ luminosity of nebulae associated with HzRGs \citep{vanojik1997} and type-II AGNs \citep{ao2017}. 
The radio luminosity of HLock01-LAB is much weaker, more than two orders of magnitude, than in HzRGs harboring luminous nebula. In fact, HLock01-LAB is not detected in the Faint Images of the Radio Sky at Twenty-cm (FIRST) radio catalog \citep[flux density limit of $\simeq 0.9$ mJy;][]{becker1995}, highlighting the importance of deep radio data in characterizing the physical mechanisms that power Ly$\alpha$ nebulae \citep[e.g.,][]{ao2017}.

Secondly, extended Ly$\alpha$ emission has been found beyond the radio structures in some HzRGs \citep[e.g.,][]{vanojik1997, maxfield2002, villar2002, villar2003, humphrey2008}, although the relative extension of the ionized gas and the extremities of the radio structures rarely exceeds a factor of two (with some exceptions e.g., 0943-242 in  \citealt{villar2003}, or TN J1338-1942 in \citealt{swinbank2015}).
In the case of HLock01-LAB, the Ly$\alpha$ emission extends over $\sim 110$ kpc, whereas the radio components are contained within the central $\sim 8$ kpc. Extended emission in metal lines, such as C~{\sc iv} (and C~{\sc iii}]), is also detected well beyond the radio structures suggesting that the material within the nebula is not primordial. 
Despite the low surface brightness limit of our VLA radio data, the detection of metal lines well beyond the radio structures may indicate that the AGN activity (as well as the radio jets) had to be more intense in the past than what we observe today. 

Lastly, the emission line ratios of Ly$\alpha$/C{\sc iv} and Ly$\alpha$/He{\sc ii} seen in the inner region of HLock01-LAB (up to $\simeq 64$ and $\simeq 82$, respectively) are one of the highest values measured to date (see also \citealt{borisova2016}, \citealt{arrigoni2018}, \citealt{shibuya2018}, \citealt{cantalupo2019}, and \citealt{marino2019}), and are much larger than those found in other nebulae around HzRGs \citep[showing typically Ly$\alpha$/C{\sc iv} and Ly$\alpha$/He{\sc ii} around 10, e.g.,:][see lower panel of Figure \ref{fig:4_7}]{villar2007}. 
As shown in Section \ref{line_ratios}, jet-induced shocks can explain relatively well the observed large ratios, further supported by the presence in this region of radio structures and gas with perturbed kinematics. 
However, it is still unclear why such large ratios are not seen also in other powerful HzRGs with clear signs of jet-gas interactions. 
A diversity of explanations may apply. Both increasing shock velocities and gas densities result in higher Ly$\alpha$ ratios relative to other emission lines \citep{allen2008, arrigoni2015b}. The radio source in HLock01-LAB is relatively small in comparison with other HzRGs, where sizes larger than few 10s of kpc are commonly observed. In these systems, the decelerated radio source may induce slower shocks that, in addition, propagate in a highly diluted medium well outside the host galaxy. In comparison, the shocks in HLock01-LAB may be propagating through the relatively dense ISM within and near the galaxy. Another relevant aspect is that Ly$\alpha$ is often strongly absorbed in HzRGs, especially in systems with strong signs of jet-gas interactions \citep{vanojik1996}. As these authors proposed, this could be a consequence of the rich cluster environment they lie in.  Although absorption is also present in  HLock01-LAB \citep[see Figure \ref{fig:4_1} or][]{marques2018}, the main effect on the Ly$\alpha$ profile is to distort the shape of the blue wing although not diminishing its flux noticeably.

\section{Summary and Conclusions}\label{sec:5}

This paper has presented the discovery and first analysis of a luminous Ly$\alpha$ nebula at $z=3.326$. HLock01-LAB was discovered close in projection ($\simeq 15^{\prime \prime}$ SW), but physically unrelated, to the gravitationally lensed system HLock01 at $z=2.96$. We have used OSIRIS on the GTC to image the Ly$\alpha$ emission with SHARDS medium-band filters and secure a rest-frame UV spectrum of the nebula, covering several UV emission lines, such as Ly$\alpha$, C~{\sc iv}, He~{\sc ii}, and C~{\sc iii}]. 
From the analysis of these data together with other existing observations covering a wide spectral range, we arrive at the following main results:

\begin{enumerate}
    \item HLock01-LAB has a total Ly$\alpha$ luminosity $L_{\rm Ly\alpha} = (6.4 \pm 0.1) \times 10^{44}$ erg s$^{-1}$, being one of the most luminous nebulae at high redshift. 
    The nebula presents an elongated morphology and extends over $\simeq 110$ kpc. Emission in C~{\sc iv} and He~{\sc ii} are also detected over a similar extension, but at much fainter flux levels.
    \item The peak of the Ly$\alpha$ emission lies very close ($\simeq 4.6$~kpc) to a central and compact galaxy (source ``A'') whose spectrum shows C~{\sc iv}, He~{\sc ii}, and C~{\sc iii}] nebular emission characteristic of a type-II AGN. We used the non-resonant He~{\sc ii} line to derive the systemic redshift $z = 3.326 \pm 0.002$.
    \item Two faint radio sources are seen on both sides of the central galaxy with a projected separation of $\simeq 15$ kpc. The non detection of a far-IR counterpart yields a radio excess $q_{\rm FIR} < 0.92$, 
    much lower than those values measured in star-forming galaxies. This implies that the radio emission is due to the AGN, rather than star formation.
    Nevertheless, the continuum emission at short wavelengths, from optical to 5.8~$\mu$m, is likely dominated by stellar emission of the host galaxy, for which we derive a stellar mass $M_{*} \simeq 2.3 \times 10^{11}$~M$_{\odot}$. 
    \item The ionized gas shows perturbed kinematics almost exclusively in the inner region between the radio structures, with $\rm FWHM > 1000$~km~s$^{-1}$, likely as a consequence of jet-gas interactions. In the outer regions of the nebula, the ionized gas presents more quiescent kinematics with line FWHM $ \lesssim 650$~km~s$^{-1}$. 
    \item Our data suggest jet-induced shocks, additional to AGN photoionization, as powering mechanisms of the Ly$\alpha$ emission. For the whole nebula, line ratios using C~{\sc iv}, He~{\sc ii}, and C~{\sc iii]} emission lines show that the gas is being photoionized by the type-II AGN. However, at the center of the nebula we find extreme line ratios of Ly$\alpha$/C{\sc iv} $\sim 50$ and Ly$\alpha$/He{\sc ii} $\sim 80$, one of the highest values measured to date, and well above the standard values of photoionization models. Jet-induced shocks are likely responsible for the Ly$\alpha$ enhancement in the center of the nebula, further supported by the presence of radio structures and perturbed kinematics in this region. 
\end{enumerate}

%\noindent
In summary, many of the properties of HLock01-LAB are broadly similar to those found in other nebulae around powerful HzRGs, yet many others have been not seen before. In particular, the large Ly$\alpha$/C{\sc iv} and Ly$\alpha$/He{\sc ii} emission line ratios observed in the inner region of HLock01-LAB, likely as a consequence of the increase of the electronic temperature from jet-induced shocks, have not been seen before in any other nebula with %clear signs of 
observational evidence of jet-gas interactions, such as those frequently associated with high redshift radio galaxies. HLock01-LAB offers, therefore, the opportunity to investigate the excitation conditions of the gas due to high-speed shocks and the underlying cooling and feedback processes. Deep and high-spectral resolution integral field spectroscopy are needed to investigate in much more detail the kinematic, ionization and morphological properties of HLock01-LAB.

\begin{acknowledgements}

We would like to thank the anonymous referee for their suggestions which significantly improved the clarity of this paper.
Based on observations made with the Gran Telescopio Canarias (GTC) and with the William Herschel Telescope (WHT), installed in the Spanish Observatorio del Roque de los Muchachos of the Instituto de Astrofísica de Canarias, in the island of La Palma. We thank the GTC and WHT staff for their help with the observations. 
R.M.C. would like to thank Claudio Dalla~Vecchia, Pablo Pérez-González, and Rosa González-Delgado for useful discussions.
R.M.C. acknowledges Fundación La Caixa for the financial support received in the form of a Ph.D. contract. R.M.C., I.P.F., L.C., P.M.N., and C.J.A. acknowledge support from the Spanish Ministerio de Ciencia, Innovación y Universidades (MICINN) under grant numbers ESP2015-65597-C4-4-R, ESP2017-86852-C4-2-R, and ESP2017-83197. 
D.R. acknowledges support from the National Science Foundation under grant number AST-1614213. 
J.L.W. acknowledges support from an STFC Ernest Rutherford Fellowship (ST/P004784/2).

\end{acknowledgements}

\bibliographystyle{aa}
%\bibliography{adssample_v2}

\end{document}